\documentclass[reprint, superscriptaddress, amsmath, amssymb, aps]{revtex4-2}

\usepackage{graphicx}
\usepackage{dcolumn}
\usepackage{bm}
\usepackage{xcolor}
\usepackage{mathtools}
\usepackage{physics}
\usepackage[normalem]{ulem}
\usepackage{braket}

\usepackage{hyperref} 
\hypersetup{colorlinks=true, citecolor=blue}

\begin{document}

\title{Use of Transmission and Reflection Complex Time Delays to Reveal Scattering Matrix Poles and Zeros: Example of the Ring Graph}

\author{Lei Chen}
 \email{LChen95@umd.edu}
 \affiliation{Maryland Quantum Materials Center, Department of Physics, University of Maryland, College Park, Maryland 20742, USA}
 \affiliation{Department of Electrical and Computer Engineering, University of Maryland, College Park, Maryland 20742, USA}

\author{Steven M. Anlage}
 \email{anlage@umd.edu}
 \affiliation{Maryland Quantum Materials Center, Department of Physics, University of Maryland, College Park, Maryland 20742, USA}
 \affiliation{Department of Electrical and Computer Engineering, University of Maryland, College Park, Maryland 20742, USA}

\date{\today}
 
\begin{abstract}
We identify the poles and zeros of the scattering matrix of a simple quantum graph by means of systematic measurement and analysis of Wigner, transmission, and reflection complex time delays.  We examine the  ring graph because it displays both shape and Feshbach resonances, the latter of which arises from an embedded eigenstate on the real frequency axis.  Our analysis provides a unified understanding of the so-called shape, Feshbach, electromagnetically-induced transparency, and Fano resonances, on the basis of the distribution of poles and zeros of the scattering matrix in the complex frequency plane.  It also provides a first-principles understanding of sharp resonant scattering features, and associated large time delay, in a variety of practical devices, including photonic microring resonators, microwave ring resonators, and mesoscopic ring-shaped conductor devices.  Our analysis is the first use of reflection time difference, as well as the first comprehensive use of complex time delay, to analyze experimental scattering data.
\end{abstract}

\maketitle

\section{Introduction}
\label{sec:Intro}

We are concerned with the general scattering properties of complex systems connected to the outside world through a finite number  of ports or channels.  The systems of interest have a closed counterpart, described by a Hamiltonian $H$, that has a spectrum of modes.  Excitations can be introduced to, or removed from, the interaction zone of the scattering system by means of the $M$ ports or channels.  The scattering matrix $S$ relates a vector of incoming (complex) waves $\ket{\psi_{in}}$ on the channels to the outgoing waves $\ket{\psi_{out}}$ on the same channels as $\ket{\psi_{out}}=S\ket{\psi_{in}}$.  The scattering matrix is a complex function of energy (or equivalently frequency) of the waves, and contains all the information about the scattering properties of the system \cite{Agassi75,Weid92,MRW2010,FSav11}.
  
Lately, there has been renewed interest in the properties of the scattering matrix in the complex frequency plane \cite{Kras19}.  This landscape is decorated with the poles and zeros of the scattering matrix, most of which lie off of the real frequency axis.  Identifying the locations of these features gives tremendous insight into the scattering properties of the system, and the movement of these features in the complex plane as the system is perturbed is also of great interest.  Knowledge of pole/zero information has practical application in the design of microwave circuits \cite{Tem77}, microwave bandpass filters \cite{Tsu02}, (where uniformity of transmission time delay is critical \cite{Gao09}), transmission through mesoscopic structures \cite{Porod93}, and the creation of embedded eigenstates \cite{EEMIT13,Kras19,Sako20}, among many other examples. Knowledge of the $S$-matrix singularities in the complex plane allows one to create coherent virtual absorption through excitation of an off-the-real-axis zero \cite{Bar17}, or virtual gain through the excitation of an off-the-real-axis pole \cite{LiVPT20}.  There is also interest in finding the non-trivial zeros of the Riemann zeta function by mapping them onto the zeros of the scattering amplitude of a quantum scattering system \cite{Rem21}. Perturbing a given system and bringing a scattering zero to the real axis enables coherent perfect absorption of all excitations incident on the scattering system \cite{CPA,Baran17,Chen2020}.  Engineering the collision of zeros and poles to create new types of scattering singularities is also of interest for applications such as sensing \cite{Mont14,Kras19,Swee20,Genack21,Genack22}.

In unitary (flux conserving) scattering systems, time delay is a real quantity measuring the time an injected excitation resides in the interaction zone before escaping through the ports \cite{Wigner55,Smith60}.  This is a well-studied quantity in the chaotic wave scattering literature, and it's statistical properties have been extensively investigated \cite{Doron90,Lehmann95,FyodSomm96,Fyodorov97,Fyodorov97a,Genack99,Brouwer99,SFS01,Texier16,Uzy17,Uzy18,Grabsch20}.  Recently, a complex generalization of time delay that applies to sub-unitary scattering systems was introduced, and this quantity turns out to be much richer than its lossless counterpart \cite{Chen2021gen,Hougne20,Chen21}.  It has been demonstrated that complex Wigner-Smith time delay is sensitive to the locations and statistics of the poles and zeros of the full scattering matrix.  One of the goals of this paper is to extend the use of complex Wigner-Smith time delay ($\tau_W$, the sum of all partial time delays) to the transmission ($\tau_T$), reflection ($\tau_R^{(1)}, \tau_R^{(2)}$, ...), and reflection time differences ($\tau_R^{(1)}-\tau_R^{(2)}$, etc.)\cite{Fyo2019,OsmanFyo20} of arbitrary multiport scattering systems.  (Note that $\tau_T$ and $\tau_R$ are complex, even for unitary scattering systems.)  This in turn yields new information about the poles and zeros of the reflection and transmission sub-matrices of $S$.  One additional novelty of our approach is the explicit inclusion of uniform attenuation in the description of the scattering system, a feature that is neglected in many other treatments of time delay, as well as treatments of scattering matrix poles and zeros.

\begin{figure}[ht]
\includegraphics[width=86mm]{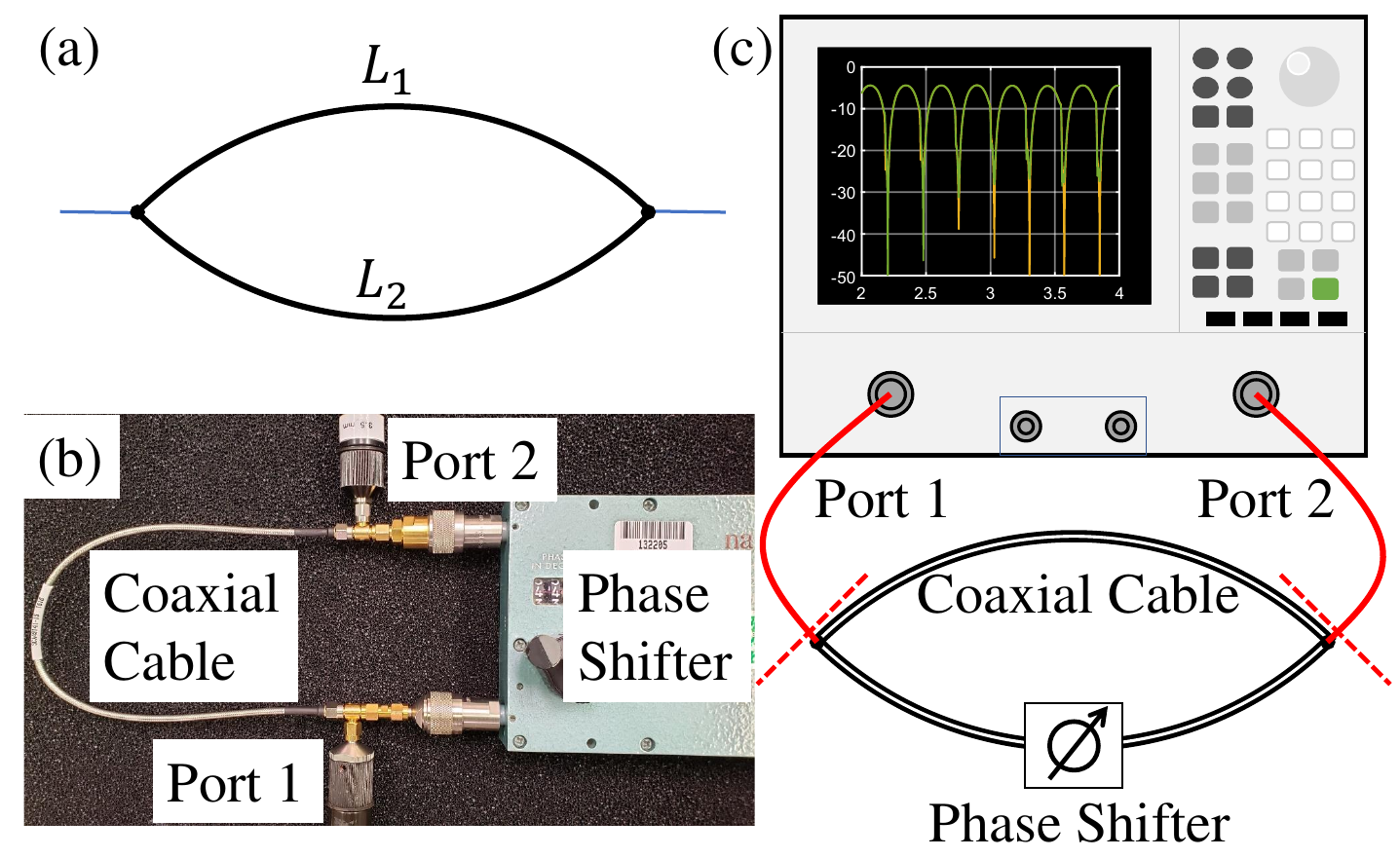}
\caption{(a) Schematic diagram of a generic ring graph connected to two infinite leads. The two bonds have length $L_1$ and $L_2$. (b) shows the picture of the experimental microwave ring graph, where a coaxial cable and a coaxial microwave phase shifter are used as the two bonds. (c) shows a schematic of the experimental setup with the microwave network analyzer included. The two dashed red lines indicate the calibration plane for the $2\times 2$ $S$-matrix measurement.}
\label{schematic}
\end{figure}

Here our attention is fixed on a simple, but remarkably important, scattering system, namely the quantum ring graph.  In this context, a graph is a network of one-dimensional bonds (transmission lines) that meet at nodes.  One can solve the Schrodinger equation for waves propagating on the bonds of metric graphs, and enforce boundary conditions at the nodes \cite{Kottos97,Kottos99,Gnu06}.  The result is a closed system in which complicated interference of waves propagating on the bonds and meeting at the nodes gives rise to a discrete set of eigenmodes.  Connecting this graph to $M$ ports (infinitely long leads) creates the scattering system of interest to us here \cite{Kottos00,Kottos03,Hul04,Hul05,Lawn08,awniczak2010}.  The ring graph, consisting of just two bonds connecting the same two nodes, which in turn are connected to $M=2$ ports (see Fig. \ref{schematic}(a)), is a ubiquitous and important scattering system.  It appears in many guises in different fields, but there is no unified treatment of its scattering properties, particularly with regard to time delay, to our knowledge.  Among other things, it forms the basis of non-reciprocal Aharonov-Bohm mesoscopic devices, as well as various types of superconducting quantum interference devices.  The scattering properties of ring graphs have been studied theoretically by a number of groups for their embedded eigenstates \cite{Exner10,Waltner2013}, and for conditions of perfect transmission \cite{Drink19,Drink20}.

Ring graphs with circumference $\Sigma$ that are on the order of the wavelength or longer, are utilized as resonators in several areas of research and applications.  Such resonators can display very narrow spectral features, which are accompanied by large time delays.  Ring resonators very elegantly and simply illustrate several different types of resonances which are known by a variety of names, including: shape modes, Feshbach modes \cite{Fesh58,Waltner2013,Chil21}, Fano modes \cite{Fano61}, electromagnetically-induced transparency (EIT) modes \cite{EIT05}, topological resonances \cite{TopoR13,Bulg17,Lawn21}, bound states in the continuum \cite{Marin08,EEMIT13,ZhenBIC14,BICRev16,Doel18}, quasinormal modes \cite{Ching98,Kris20}, etc.  Here we use the shape/Feshbach terminology to discuss the modes, but our results apply to ring graphs in all contexts.  To illustrate the ubiquity and importance of the ring graph, we next discuss some of the diverse manifestations and properties of this simple graph.

Fano resonances have been studied by many authors in the context of quantum transport through graph-like structures \cite{Porod93,Miro10,Huang15}.  The Fano resonance arises from the constructive and destructive interference of a narrow discrete resonance (typically a bound state of the closed system) with a broad spectral line or continuum excitation, thus creating two scattering channels \cite{Shao94,Luk10}.  The interference of these two channels gives rise to the celebrated Fano resonance profile \cite{Fano61,Miro10}.

EIT is a quantum phenomenon that arises from interference between transitions taking place between multiple states \cite{EIT05}.  It has a classical analog that can be realized in a wide variety of coupled oscillator scenarios \cite{Alzar02}.  For example, an EIT/Fano resonance feature was proposed for a generic resonator coupled to an optical transmission line \cite{Fan02}.  EIT phenomena have also been created through metamaterial realizations in which a strongly coupled (bright resonator) and weakly coupled (dark resonator) oscillator are brought into interference to completely cancel transmission, and at the same time create `slow light' (enhanced transmission time delay), all at one wavelength \cite{Fed07,Pap08,Kurter11}.  \\


In terms of applications, ring resonators have been employed in microwave circuit devices for many years \cite{Trought68,KaiRR04}.  It was recognized that pairs of nearly degenerate modes exist in this structure and their interference could be used to advantage \cite{Wolff72,KaiRR04}.  Microstrip ring resonators are routinely created with intentional defects or stubs in one arm, or are coupled asymmetrically, to create interference of the nearly degenerate modes \cite{KaiRR04}.  

EIT-like resonant features have been created in optical microring resonators coupled to transmission lines by a number of groups.  A classical analog of EIT was demonstrated with two photonic ring resonators coupled to optical fibers \cite{EITOp06}.  A set of two coupled microspheres, acting as ring resonators, showed the classical analog of EIT for light, and demonstrated large transmission time delay \cite{Tot07}.  An integrated optical waveguide realization of the ring graph, with one arm hosting a variable delay element, has been used to create “EIT dips” with associated large transmission delay \cite{Bog17}.  Other work has used a pair of Silicon microring photonic resonators to create a non-reciprocal diode effect for light (1630 nm) by exploiting a Fano resonance and nonlinearity \cite{LiFano12}. \\

Mesoscopic ring graph structures made of metals and semiconductors have been studied extensively for evidence of electron interference in their transport properties \cite{But84,Kowal90,Datta95}.  Much of this work is focused on rings immersed in a magnetic field and showing quantum interference properties arising from the Aharonov-Bohm (AB) effect \cite{AB59,Webb85}.  Aharonov-Bohm rings with a localized trapping site in one arm have been proposed to generate non-reciprocal transmission time delay \cite{Mann21}, and asymmetric transport \cite{Bred21}.  \\

Finally, superconducting quantum interference devices (SQUIDs) are based on a loop graph structure that supports a complex superconducting order parameter.  The closed loop structure creates a quantization condition for the magnetic fluxoid, and the addition of one or more Josephson junctions to the ring bonds, along with the addition of two leads, creates a sensitive magnetic flux to voltage transducer known as a dc SQUID \cite{Jak64,Silver67,Tinkham}. 

The purpose of this paper is to apply the complex time delay approach to experimental data on a microwave realization of the ring graph with the goal of identifying the complete set of scattering poles, as well as scattering, transmission and reflection zeros, of the graph.  With this information we are able to thoroughly characterize the scattering properties of this system, and at the same time establish a basis that unifies the many disparate approaches to describing the scattering properties of this remarkable graph.\\  

The outline of this paper is as follows.  In Section \ref{sec:compTDs}, we present expressions for the complex times delays in terms of singularities of the scattering matrix.  In Section \ref{sec:RingGraph}, we discuss the properties of the ring graph, including the predicted locations of its poles and zeros in the complex plane.  Section \ref{sec:Exp} presents our experiment on the microwave realization of the ring graph and measurements of the scattering matrix, and Section \ref{sec:DataAnalysis} presents the complex time delays extracted from the measured $S$-matrix as a function of frequency, as well as fits to reveal the locations of the scattering singularities.  Section \ref{sec:detSAnalysis} uses the results from Section \ref{sec:DataAnalysis} to reconstruct $\det[S]$ over the entire complex frequency plane. This is followed by discussion of all the results in Section \ref{sec:Discuss}, and then conclusions in Section \ref{sec:Conclusions}.

\section{Complex Time Delays and Scattering Poles and Zeros}
\label{sec:compTDs}

A useful theoretical framework for the complex time delay analysis is the so called effective Hamiltonian formalism for wave-chaotic scattering \cite{SokZel89,Fyodorov97,FSav11,kuhl13,Schomerus2015}. It starts with defining an $N \times N$ self-adjoint matrix Hamiltonian $H$ whose real eigenvalues are associated with eigenfrequencies of the closed system. Further defining $W$ to be an $N \times M$ matrix of coupling elements between the $N$ modes of $H$ and the $M$ scattering channels, one can build the unitary $M\times M$ scattering matrix $S(E)$ in the form:
\begin{align}
    \label{Smodel}
    S(E)=1_M - 2\pi i W^{\dag} \frac{1}{E-H+i\Gamma_W} W,
\end{align}
where we defined $\Gamma_W = \pi WW^{\dag}$. Note that in this approach the $S$-matrix poles $\mathcal{E}_n = E_n -i\Gamma_n$ (with $\Gamma_n>0$) are complex eigenvalues of the non-Hermitian effective Hamiltonian matrix ${\cal H}_\text{eff}=H-i\Gamma_W\ne {\cal H}_\text{eff}^{\dagger}$.

A standard way of incorporating the uniform absorption with strength $\eta$ is to replace $E\to E+i \eta$ in the $S$ matrix definition. Such an $S$-matrix becomes subunitary and we denote $S(E + i\eta) \coloneqq S_{\eta}(E)$. The determinant of $S_{\eta}(E)$ is then given by
\begin{align}
    \label{detS}
    \det S_{\eta}(E) &\coloneqq \det S(E+i\eta) \\
    \label{detSmatrix}
    &=\frac{\det [E-H+i(\eta  - \Gamma_{W})]}
    {\det [E-H+i(\eta  + \Gamma_{W})]} \\
    \label{detSmodel}
    &=\prod_{n=1}^{N} \frac{E+i\eta - z_n}{E+i\eta - \mathcal{E}_n},
\end{align}
where Eq. (\ref{detSmatrix}) follows from Eq. (\ref{Smodel}), and Eq. (\ref{detSmodel}) expresses the determinants in terms of the eigenvalues of the non-Hermitian matrices involved.  Here the $S$-matrix zeros $z_n$ are complex eigenvalues of the non-Hermitian matrix ${\cal H}_\text{eff}^{\dagger} = H+i\Gamma_W$, i.e. $z_n = \mathcal{E}_n^*$.  

Using the above expression, the Wigner-Smith (which we shall abbreviate as Wigner) time delay can be very naturally extended to scattering systems with uniform absorption as suggested in \cite{Chen2021gen} by defining:
\begin{widetext}
\begin{align}
    \label{WTD}
    \tau_W(E;\eta) &\coloneqq \frac{-i}{M} \frac{\partial}{\partial E} \log \det S(E+i\eta) \\
    \label{CWTD}
    &= \text{Re}\ \tau_W(E;\eta) + i\text{Im}\ \tau_W(E;\eta), \\
    \label{rWTD}
    \text{Re}\ \tau_W(E;\eta) &= \frac{1}{M} \sum_{n=1}^{N} \left[ \frac{\Gamma_n - \eta}{(E-E_n)^2 + (\Gamma_n - \eta)^2} + \frac{\Gamma_n + \eta}{(E-E_n)^2 + (\Gamma_n + \eta)^2} \right], \\
    \label{iWTD}
    \text{Im}\ \tau_W(E;\eta) &= -\frac{1}{M} \sum_{n=1}^{N} \left[ \frac{E - E_n}{(E-E_n)^2 + (\Gamma_n - \eta)^2} - \frac{E - E_n}{(E-E_n)^2 + (\Gamma_n + \eta)^2} \right].
\end{align}
\end{widetext}

We note that the complex Wigner time delay is a sum of Lorentzians whose properties depend on the poles and zeros of the full scattering matrix, as well as the uniform absorption.  Prior work has shown that Eqs. (\ref{rWTD}) and (\ref{iWTD}) provide an excellent description of the experimental complex time delay for isolated modes of a lossy tetrahedral microwave graph \cite{Chen2021gen}.  The statistical properties of complex time delay in an ensemble of tetrahedral graphs are also in agreement with those based on Eqs. (\ref{rWTD}) and (\ref{iWTD}) and the random matrix theory predictions for the distribution of $\Gamma_n$ \cite{Chen21}.

We can define the scattering matrix as $S=\begin{pmatrix}
R & T'\\
T & R'
\end{pmatrix}$ in terms of the reflection sub-matrix $R$ and transmission sub-matrix $T$ \cite{Fish81,Rot17,Genack21,Genack22}. For a system with uniform absorption, the determinant of the transmission sub-matrix can be written as:
\begin{align}
    \label{dettmatrix}
    \det T_{\eta}(E) = (-2\pi i)^M \frac{\det (E-H+i\eta) \det (W_2^{\dagger} \frac{1}{E-H+i\eta} W_1)}{\det [E-H+i(\eta  + \Gamma_{W})]},
\end{align}
where the coupling matrix $W = [W_1\ W_2]$, is decomposed into its port-specific $N\times M$ coupling matrices $W_{1/2}$.
We can extend the transmission time delay \cite{Genack21} into a complex quantity: 
\begin{widetext}
\begin{align}
    \label{TTD}
    \tau_T(E;\eta) &\coloneqq -i \frac{\partial}{\partial E} \log \det T(E+i\eta) \\
    \label{CTTD}
    &= \text{Re}\ \tau_T(E;\eta) + i\text{Im}\ \tau_T(E;\eta), \\
    \label{rTTD}
    \text{Re}\ \tau_T(E;\eta) &= \sum_{n=1}^{N-M} \frac{\text{Im}\ t_n - \eta}{(E-\text{Re}\ t_n)^2 + (\text{Im}\ t_n - \eta)^2} + \sum_{n=1}^{N} \frac{\Gamma_n + \eta}{(E-E_n)^2 + (\Gamma_n + \eta)^2}, \\
    \label{iTTD}
    \text{Im}\ \tau_T(E;\eta) &= - \left\{ \sum_{n=1}^{N-M} \frac{E-\text{Re}\ t_n}{(E-\text{Re}\ t_n)^2 + (\text{Im}\ t_n - \eta)^2} - \sum_{n=1}^{N} \frac{E - E_n}{(E-E_n)^2 + (\Gamma_n + \eta)^2} \right\}.
\end{align}
\end{widetext}

Here $t_n = \text{Re}\ t_n + i\text{Im}\ t_n$ denote the complex zeros of $\det (T)$, while $\mathcal{E}_n = E_n -i\Gamma_n$ are the same poles defined in Eq. (\ref{detSmodel}).  Note in Eqs. (\ref{rTTD}) and (\ref{iTTD}) that the number of zero-related terms is smaller than the number of pole-related terms \cite{Genack21}. 

Recent interest in the zeros of the $S$-matrix in the complex energy plane has motivated the use of the Heidelberg model to introduce the concept of reflection time delays \cite{Fyo2019,OsmanFyo20}.  To begin with, consider the special case of a two-channel ($M=2$) flux-conserving scattering system which can be described by the $2\times2$ unitary scattering matrix:
\begin{align}
    \label{Rmatrix}
    S(E) &= 
    \begin{pmatrix}
    R_1(E) & t_{12}(E) \\
    t_{21}(E) & R_2(E)
    \end{pmatrix}. 
\end{align}
The two reflection elements $R_{1,2}(E)$ at both channels may have zeros $r_n$ in the complex energy plane. 

In the presence of uniform absorption strength $\eta$, the full scattering matrix $S$ becomes sub-unitary, and $|R_1(E+i\eta)| \neq |R_2(E+i\eta)|$ in general. In that case, the reflection element $R_1(E+i\eta)$ at channel 1 can be written in a similar form to the $\det S_{\eta}$ and $\det T_{\eta}$ formalism:
\begin{align}
    \label{R1matrix}
    R_1(E+i\eta) &= \frac{\det [E-H+i(\eta - \Gamma_{W}^{(1)} + \Gamma_{W}^{(2)})]}
    {\det [E-H+i(\eta + \Gamma_{W})]} \\
    \label{R1model}
    &= \prod_{n=1}^{N} \frac{E+i\eta - r_n}{E+i\eta - \mathcal{E}_n},
\end{align}
where $\Gamma_W=\Gamma_{W}^{(1)} + \Gamma_{W}^{(2)}$, and $r_n=u_n+iv_n$ are the positions of reflection zeros, which are the complex eigenvalues of $H+i(\Gamma_{W}^{(1)} - \Gamma_{W}^{(2)})$. Similarly, the reflection element $R_2(E+i\eta)$ at channel 2 can be written as
\begin{align}
    \label{R2matrix}
    R_2(E+i\eta) &= \frac{\det [E-H+i(\eta - \Gamma_{W}^{(2)} + \Gamma_{W}^{(1)})]}
    {\det [E-H+i(\eta + \Gamma_{W})]} \\
    \label{R2model}
    &= \prod_{n=1}^{N} \frac{E+i\eta - r_n^{*}}{E+i\eta - \mathcal{E}_n}.
\end{align}
Thus, the reflection time delays in uniformly absorbing systems are introduced as
\begin{align}
    \label{RTD1}
    \tau_R^{(1)}(E;\eta) \coloneqq -i \frac{\partial}{\partial E} \log R_{1}(E+i\eta) 
\end{align}
and
\begin{align}
    \label{RTD2}
    \tau_R^{(2)}(E;\eta) \coloneqq -i \frac{\partial}{\partial E} \log R_{2}(E+i\eta).
\end{align}

In full analogy with the complex Wigner time delay model, the complex reflection time delay for channel 1, $\tau_R^{(1)}(E;\eta)$, is given by
\begin{widetext}
\begin{align}
    \label{rRTD1}
    \text{Re}\ \tau_R^{(1)}(E;\eta) &= \sum_{n=1}^{N} \left[ \frac{v_n - \eta}{(E-u_n)^2 + (v_n - \eta)^2} + \frac{\Gamma_n + \eta}{(E-E_n)^2 + (\Gamma_n + \eta)^2} \right], \\
    \label{iRTD1}
    \text{Im}\ \tau_R^{(1)}(E;\eta) &= - \sum_{n=1}^{N} \left[ \frac{E-u_n}{(E-u_n)^2 + (v_n - \eta)^2} - \frac{E-E_n}{(E-E_n)^2 + (\Gamma_n + \eta)^2} \right].
\end{align}
\end{widetext}
Similarly, we also have the complex reflection time delay for channel 2, $\tau_R^{(2)}(E;\eta)$:
\begin{widetext}
\begin{align}
    \label{rRTD2}
    \text{Re}\ \tau_R^{(2)}(E;\eta) &= \sum_{n=1}^{N} \left[ \frac{-v_n - \eta}{(E-u_n)^2 + (v_n + \eta)^2} + \frac{\Gamma_n + \eta}{(E-E_n)^2 + (\Gamma_n + \eta)^2} \right], \\
    \label{iRTD2}
    \text{Im}\ \tau_R^{(2)}(E;\eta) &= - \sum_{n=1}^{N} \left[ \frac{E-u_n}{(E-u_n)^2 + (v_n + \eta)^2} - \frac{E-E_n}{(E-E_n)^2 + (\Gamma_n + \eta)^2} \right].
\end{align}
\end{widetext}

Notice that the two reflection time delays share the same terms arising from the $S$-matrix poles, thus another useful quantity, the complex reflection time difference, can be defined as $\delta \mathcal{T}_R(E;\eta) := \tau_R^{(1)}(E;\eta) - \tau_R^{(2)}(E;\eta)$ \cite{Fyo2019,OsmanFyo20}:
\begin{widetext}
\begin{align}
    \label{rRTDD}
    \text{Re}\ \delta \mathcal{T}_R(E;\eta) = \text{Re}\ \tau_R^{(1)}(E;\eta) - \text{Re}\ \tau_R^{(2)}(E;\eta) &= \sum_{n=1}^{N} \left[ \frac{v_n - \eta}{(E-u_n)^2 + (v_n - \eta)^2} + \frac{v_n + \eta}{(E-u_n)^2 + (v_n + \eta)^2} \right], \\
    \label{iRTDD}
    \text{Im}\ \delta \mathcal{T}_R(E;\eta) = \text{Im}\ \tau_R^{(1)}(E;\eta) - \text{Im}\ \tau_R^{(2)}(E;\eta) &= - \sum_{n=1}^{N} \left[ \frac{E-u_n}{(E-u_n)^2 + (v_n - \eta)^2} - \frac{E-u_n}{(E-u_n)^2 + (v_n + \eta)^2} \right].
\end{align}
\end{widetext}
The reflection time difference is determined solely by the position of the reflection zeros, and has no contribution from the poles.

Our approach to defining and utilizing multiple types of complex time delay overcomes a number of issues with prior treatments.  First, we treat poles and zeros on an equal footing, as both contribute significantly to the complex time delay.  Secondly, the imaginary part of the time delay provides redundant, but nevertheless useful, information about the pole/zero locations.  The imaginary part has one advantage over the real part in terms of fitting to find pole and zero locations: the imaginary part changes sign at each singularity, leading to smaller tails at the locations of nearby singularities.  This is particularly useful for systems with a dense set of modes.  In all examples below, we fit \underline{both} quantities simultaneously using a single set of fitting parameters.  Finally, our approach directly includes the effect of \textit{uniform} loss, frequently ignored in most prior treatments of time delay.  Note that we have previously examined the effects of varying \textit{lumped} loss on the complex Wigner time delay \cite{Chen2021gen}, and observed the resulting independent motion of the poles and zeros in the complex plane (i.e. violating the condition that $z_n = \mathcal{E}_n^*$, for example) \cite{Li2017,Fyodorov2017,Fyo2019,OsmanFyo20}.\\

We note in passing that the use of complex time delay will enhance the study of scattering phenomena governed by pole/zero distributions.  We have demonstrated this in the context of CPA \cite{Chen2021gen,Frazier20}, and the generation of ``cold spots", in complex scattering systems \cite{Frazier20}.  Further opportunities await for the generalized Wigner-Smith operator \cite{Horodynski20}, and for the generation of ``slow light."\\

Finally, we note that although the Wigner-Smith time delay is purely real for unitary scattering systems, the reflection and transmission time delays are always complex, due to the fact that they are derived from sub-unitary parts of the full $S$-matrix.  Thus a proper treatment of these delays must take into account their complex nature, even in the flux-conserving limit. \\

\section{The Ring Graph}
\label{sec:RingGraph}

Ring graph structures have appeared in quantum graph studies, mesoscopic devices, microwave ring resonators, optical micro-ring resonators, and superconducting quantum interference devices.  It is a generic and important structure for wave systems because it is a simple way to introduce wave interference phenomena in a controlled manner.\\

As shown in the schematic diagram in Fig. \ref{schematic}(a), the ring graph has two bonds, of lengths $L_1$ and $L_2$, connecting two nodes.  We assume that the bonds of the graph support travelling waves in both directions, with identical propagation and loss characteristics.  The nodes are also connected to infinite leads (ports).  Coupling between the leads and ring graph is provided by means of a 3-way tee junction with ideal scattering matrix 
\[ S_\text{tee} = \begin{pmatrix} -1/3 & 2/3 & 2/3 \\ 2/3 & -1/3 & 2/3 \\ 2/3 & 2/3 & -1/3 \end{pmatrix}. \]
We shall investigate the $M=2$ scattering matrix $S$ between the left lead and the right lead in Fig. \ref{schematic}(a).
Two cases are of interest to us here: i) rationally-related bond lengths $L_1$ and $L_2$, including the case $L_1=L_2$, and ii)  irrationally-related lengths $L_1$ and $L_2$.\\


A metric ring graph with $L_1=L_2$ can support two distinct eigenmodes.  Each involves spanning the circumference of the graph $\Sigma=L_1+L_2$ with an integer number of wavelengths of the wave excitation.  One mode, which we call the shape resonance, has a maximum of the standing wave pattern at the nodes of the graph \cite{Exner10}.  The second mode has a standing wave pattern that is rotated one quarter of a wavelength relative to the first and has zero amplitude at the nodes.   Such an embedded eigenstate on a ring graph with rationally-related bond lengths can have a compact eigenfunction even though the graph extends to infinity.  In other words, the eigenmode is nonzero over most of the ring graph, but has zero amplitude at the locations of the leads, preventing the mode from extending into the leads.  This means that the the eigenvalue can be in a continuum of states, but the eigenstate can have no amplitude on the leads of the graph.   Small perturbations to the length(s) of the bond will move the pole off of the real axis and produce a narrow high-Q resonance, along with a nearby complex zero.  This is known as a Feshbach mode.\\

Waltner and Smilansky \cite{Waltner2013} have made predictions for the $S$-matrix zeros and poles for both shape and Feshbach resonances of the ring graph. In the case of a symmetrical graph (i.e. $L_1 = L_2$), or for graphs with rationally related lengths, the scattering properties of the graph show shape resonances only. The $S$-matrix poles of the shape resonances are given by 
\begin{align}
    \label{ShapePole}
    \mathcal{E}_n^{S, symm} = nc/\Sigma - i\ c\ln{3}/(\pi \Sigma), 
\end{align}
where $\Sigma = L_1 + L_2$ is the total \textit{electrical} length of the ring graph, $c$ is the speed of light in vacuum (here we specialize to the case of microwave ring graphs), and $n$ is the mode index ($n = 1, 2, 3, ...$).  The $S$-matrix zeros are simply the complex conjugates of the poles:
\begin{align}
    \label{ShapeZero}
    z_n^{S, symm} = nc/\Sigma + i\ c\ln{3}/(\pi \Sigma). 
\end{align}
The Feshbach modes are not visible in this case. 

In the case of a non-symmetrical graph (i.e. $\delta = L_1 - L_2 \neq 0$ and $L_1/L_2$ is not rational, the graph has both shape and Feshbach resonances. In the limit of $n\delta \ll \Sigma$, the $S$-matrix poles of the Feshbach resonances are given by 
\begin{align}
    \label{FeshPole}
    \mathcal{E}_n^{F, asymm} \approx nc/\Sigma - i\ (c/2\pi)[(2\pi n\delta)^2/(8{\Sigma^3})],
\end{align}
while the poles of the shape resonances become $\mathcal{E}_n^{S, asymm} \approx (nc/\Sigma + \alpha) - i\ [c\ln{3}/(\pi \Sigma) + \beta]$, where $\alpha = nc\delta^2\ln{3}/(2{\Sigma^3})$ and $\beta = (c/2\pi)[(2\ln{3})^2 - (2\pi n)^2]\delta^2/(8{\Sigma^3})$ are small changes compared to the original pole locations, Eq. (\ref{ShapePole}).  Again the $S$-matrix zeros are complex conjugates of the pole locations: $z_n^{F, asymm} = [\mathcal{E}_n^{F, asymm}]^*$ and $z_n^{S, asymm} = [\mathcal{E}_n^{S, asymm}]^*$.  These predictions will be tested in our analysis of complex time delay data below.

We note that the imaginary part of the Feshbach pole (and zero) in Eq. (\ref{FeshPole}) increases in magnitude as $(n\delta)^2$.  The Warsaw group has studied the length asymmetry ($\delta$) dependence of the lowest frequency ($n=1$) pole of the ring graph \cite{Lawn21}.  A cold atom collision experiment has observed the flow of the shape and Feshbach resonance poles as the system is perturbed \cite{Chil21}.   In contrast with earlier work, we study the dependence of the poles and zeros at two fixed bond lengths upon the mode index $n$, among other things.\\

\section{Experiment}
\label{sec:Exp}

A picture of the ring graph experimental setup is shown in Fig. \ref{schematic}(b). A 15-inch (38.1 cm) long coaxial cable is used as the fixed length bond $L_1$, while a mechanically-variable coaxial phase shifter is used as the variable length bond $L_2$.  The coaxial cable has a center conductor that is 0.036 in (0.92 mm) in diameter, a Teflon dielectric layer (with $\epsilon_r =2.1$ and $\mu_r =1$), and an outer conductor that is 0.117 in (2.98 mm) in diameter.  The center conductor is silver-plated copper-clad steel, while the outer conductor is copper-tin composite.  The electrical length of the cable is given by the product of the geometrical length and the index of refraction, $\sqrt{\epsilon_r \mu_r}$.  The phase shifter is a Model 3753B coaxial phase shifter from L3Harris Narda-MITEQ that provides up to 60 degrees of phase shift per GHz.  The measurement cables (leads) are connected to the ring graph through two Tee junctions, acting as the nodes. When the graph is symmetrical (i.e. $L_1 = L_2$), the total electrical length of the graph is $\Sigma_{symm} = 1.0993$ m.  The graph shows a mean spacing between shape modes of $\Delta f = 0.2729$ GHz, giving rise to a Heisenberg time $\tau_H = 2\pi/\Delta f$ of 23.02 ns.  We measure the scattering response from all the modes spanning the frequency range from 0 to 10 GHz, encompassing modes $n=1$ to $n=37$.

\begin{figure}[ht]
\includegraphics[width=86mm]{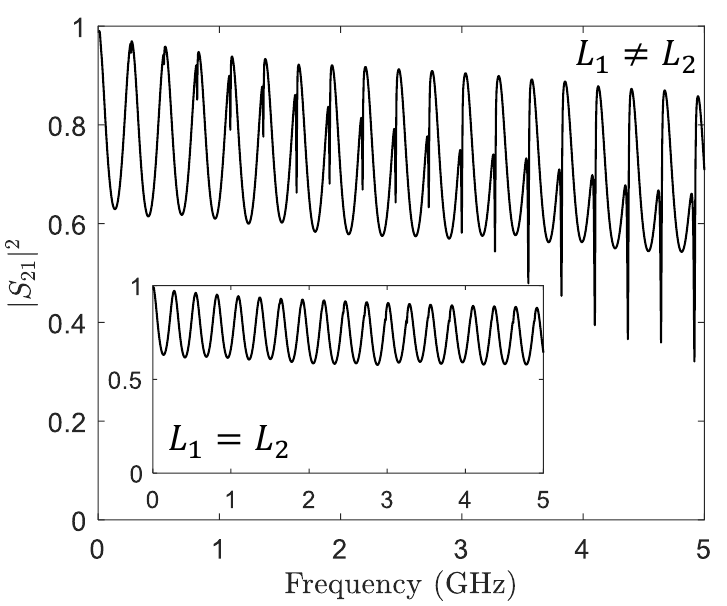}
\caption{Transmission spectrum $|S_{21}|^2$ vs. frequency measured for the first 18 modes of a microwave ring graph. Main figure shows the transmission of non-equal lengths ($L_1\neq L_2$) between the phase shifter and the coaxial cable, while the inset shows the case of equal lengths ($L_1 = L_2$). The sinusoidal wiggles come from the shape resonances, while the narrow dips come from the Feshbach resonances. Note that the data in the inset shows no narrow resonances.}
\label{Transmission}
\end{figure}

To make the graph asymmetric ($L_1 \neq L_2$) we set the phase shifter to produce $\delta = 0.577$ cm.  Thus we maintain the condition $n\delta \ll \Sigma$ up to $n=37$.

The time delay analysis involves taking frequency derivatives of the measured $S$-matrix phase and amplitude data, and this demands fine frequency resolution and careful measurement. In order to obtain high-quality data, we first conducted a careful calibration of the Agilent model N5242A microwave vector network analyzer (VNA), utilizing an intermediate frequency (IF) bandwidth of 100 Hz and a frequency step size of $84.375$ kHz (about $3\times 10^{-4}$ of the mean spacing between shape resonances).  The calibration process creates boundary conditions for the microwaves that are equivalent to the presence of the two infinite leads connected to the nodes of the ring graph.  In other words, waves exiting the system will never return.  In addition, the scattering matrix is evaluated at the plane of calibration as the ratio of ingoing and outgoing complex waves measured at that point.  The plane of calibration is at the two nodes labelled by red dashed lines in Fig. \ref{schematic}(c). We then measured the $2\times2$ $S$-matrix of the graphs with the same settings of the VNA.  By doing so, we minimize the measurement noise and acquire high resolution data.  The phase of the $S$-matrix data was unwound into a continuous variation to eliminate artificial discontinuities in time delay due to $2\pi$ phase jumps.  We also developed an algorithm for taking numerical derivatives of the experimental data utilizing variable frequency window smoothing settings. Given the number of data points in a smoothing window, we obtained the overall slope through a line fitting of all the data samples. The size of the smoothing window can be dynamically adjusted based on the variability of the phase and amplitude with frequency. All of these steps are required to generate high-quality time delay data for further analysis. Note that the numerical derivatives are taken on the raw $S$-matrix data without any normalization step or background subtraction, etc. There is no need to augment or modify the raw $S$-matrix data, as it contains all the information about the graph, including coupling, loss, and scattering singularities.

The two types of modes present in the ring graph, namely shape resonances and Feshbach resonances, are illustrated in the measured transmission $|S_{21}|^2$ vs. frequency plot shown in Fig. \ref{Transmission}.  The inset in Fig.  \ref{Transmission} shows the transmission spectrum when the two bond lengths are equal ($L_1 = L_2$).  In this case only the shape resonances appear in the scattering data.  For the main plot in Fig. \ref{Transmission}, we tuned the electrical length of the phase shifter so that the two bonds lengths are not equal ($L_1 \neq L_2$) and not rationally related. The narrow Feshbach resonances occur at lower frequencies than the shape resonances and their separation from the shape resonances grows with mode number $n$, as predicted by Eq. (\ref{FeshPole}), and demonstrated in the following analysis.

\section{Complex Time delay analysis on Ring graph data}
\label{sec:DataAnalysis}

\begin{figure}[ht]
\includegraphics[width=86mm]{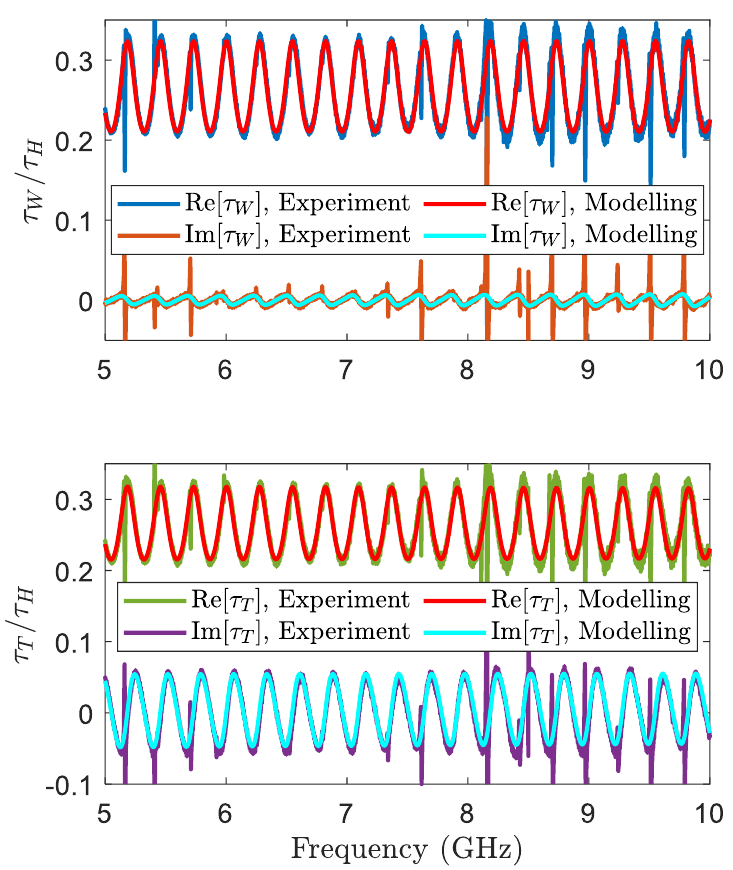}
\caption{Comparisons between the experimental data and the modelling for the complex Wigner time delay (upper plot) and for the complex transmission time delay (lower plot), both normalized by the Heisenberg time $\tau_H$, as a function of frequency for a symmetric ($L_1=L_2$) microwave ring graph. The modelling data are plotted on top of the experimental data, and are in good agreement.}
\label{Shape_WignerTransmission}
\end{figure}

\begin{figure}[ht]
\includegraphics[width=86mm]{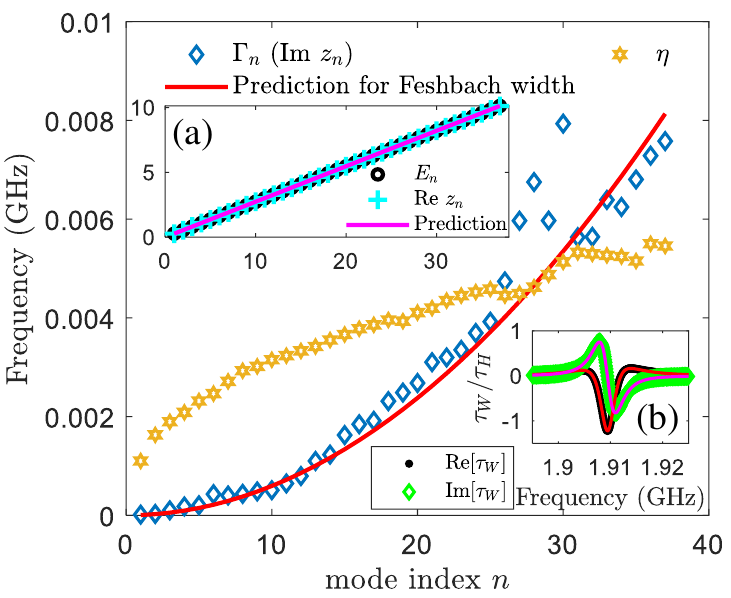}
\caption{Comparison between fitted pole location parameters ($\mathcal{E}_n = E_n -i\Gamma_n$) and predictions for multiple Feshbach modes of the asymmetric microwave ring graph ($L_1 \neq L_2$). Inset (a) shows the comparison between fitted real parts of the zeros and the poles, along with the prediction by Eq. (\ref{FeshPole}) shown as a straight purple line. Inset (b) shows such a representative fit to $\tau_W(f)$ for a single Feshbach mode ($n=7$).}
\label{Feshbach_Wigner}
\end{figure}

In the case of a symmetrical graph, we analyze the complex Wigner time delay and transmission time delay properties of the shape resonances alone. Figure \ref{Shape_WignerTransmission} shows the complex Wigner ($\tau_W$) and transmission ($\tau_T$) time delay as a function of frequency over 18 modes of the ring graph. The two time delays are calculated from the measured $S$-matrix based on Eqs. (\ref{WTD}) (Wigner) and (\ref{TTD}) (Transmission), respectively. Note that in all comparisons of data and theory we treat frequency $f$ and energy $E$ as equivalent.  We also reconstruct the two time delays based on the models from Eqs. (\ref{rWTD}) \& (\ref{iWTD}) (Wigner) and Eqs. (\ref{rTTD}) \& (\ref{iTTD}) (Transmission), using the scattering matrix poles prediction from Eq. (\ref{ShapePole}) and the zeros from Eq. (\ref{ShapeZero}). The poles are calculated based on the measured dimension (electrical length) of the ring graph, and the zeros are assumed to be the complex conjugates of the poles. The modelled complex time delays are plotted with the experimental data in Fig. \ref{Shape_WignerTransmission}, and are in good agreement. (Due to uncertainties in the lengths of the components, we adjusted $\Sigma$ slightly to precisely match the $\tau_W$ frequency dependence in Fig. \ref{Shape_WignerTransmission}.)  Note that in the complex transmission time delay modelling we use only the pole information (there are no transmission zeros in this case due the absence of an interfering mode \cite{Shao94}), while in the complex Wigner time delay modelling we use both the pole and zero information.  

We note that although the model is in very good agreement with the data in Fig. \ref{Shape_WignerTransmission} there are a number of sharp vertical features in the data that are not reproduced by the model.  Theoretical treatments of a delta function scatterer in the ring graph shows that imperfections in a symmetric graph ($L_1=L_2$) can give rise to Feshbach resonances \cite{Waltner2013,Walt14}.  We interpret the spikes seen in $\tau_W$ and $\tau_T$ as arising from impedance discontinuities in the phase shifter and its coaxial connectors, acting effectively as delta-function scatterers.  To verify this, we measured a symmetric graph made up of two identical fixed-length (15 inch) coaxial cables and found that there are no sharp vertical features in the time delays in that case.

\begin{figure}[ht]
\includegraphics[width=86mm]{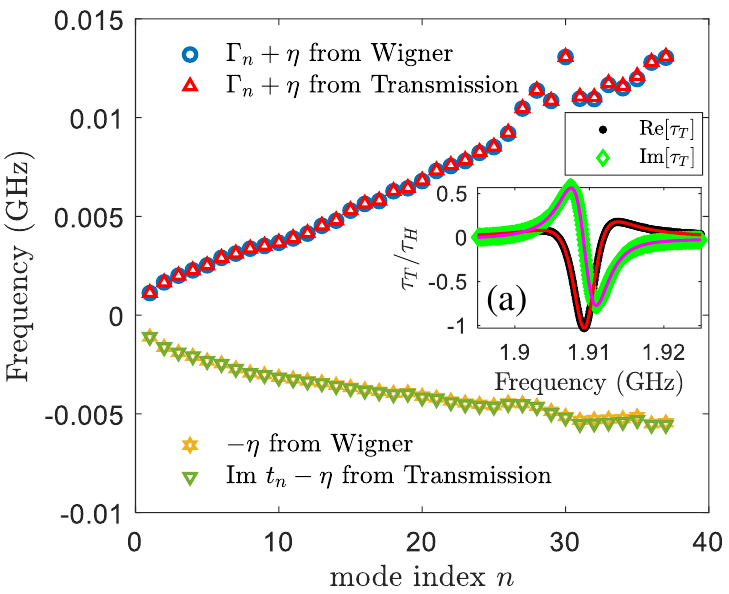}
\caption{Comparison between fitted pole location parameters ($\mathcal{E}_n = E_n -i\Gamma_n$) obtained from the complex Wigner time delay (blue circles) and the complex transmission time delay (red triangles) for Feshbach modes of the asymmetric ring graph. The lower part of the figure shows the comparison between fitted uniform attenuation ($- \eta$) obtained from the complex Wigner time delay (yellow stars) in Fig. \ref{Feshbach_Wigner} and fitted imaginary parts of the transmission zeros ($\text{Im}\ t_n - \eta$) obtained from the complex transmission time delay (green triangles) on all measured Feshbach modes. Inset (a) shows a representative fit to $\tau_T(f)$ for a single Feshbach mode ($n=7$).}
\label{Feshbach_Transmission}
\end{figure}

Next we analyze the complex Wigner time delay and transmission time delay properties for the Feshbach resonances of the ring graph. We tuned the electrical length of the phase shifter so that the two bonds lengths are not equal or rationally related (with $\delta \approx$ 0.577 cm), and a set of Feshbach resonances appear, as in Fig.  \ref{Transmission}. We followed the same procedure to calculate the complex Wigner and transmission time delay from the newly measured $S$-matrix. Note that the shape resonances are always present in the system. We first removed the effects of the shape resonances from the overall time delay data by subtracting their contributions to the time delay data. The contributions from the shape resonances are modelled in the same way as demonstrated in Fig. \ref{Shape_WignerTransmission}. ($\Sigma$ has been slightly adjusted to accommodate the length change of the ring graph system.)  We then fit the remaining complex time delay data with the model Eqs. (\ref{rWTD}) \& (\ref{iWTD}) (Wigner) and Eqs. (\ref{rTTD}) \& (\ref{iTTD}) (Transmission), for each individual Feshbach mode. Both the zero and pole locations, as well as the uniform absorption strength $\eta$, are used as fitting parameters in this process.  Note that the real and imaginary parts of each time delay are fit simultaneously with a single set of parameters.  We also constrain the zeros to be complex conjugates of the poles during the Wigner time delay fitting. One fitting example is shown in Figs. \ref{Feshbach_Wigner}(b) (Wigner) and \ref{Feshbach_Transmission}(b) (Transmission), respectively. The fitting process was repeated for all 37 modes measured, and all fits were very successful (see Appendix \ref{appendix:TZeros} for further discussion about the transmission zeros). The fit parameters for the complex zeros and poles, as well as the uniform attenuation, are plotted in Figs. \ref{Feshbach_Wigner} (Wigner) and \ref{Feshbach_Transmission} (Transmission), respectively.

\begin{figure}[ht]
\includegraphics[width=86mm]{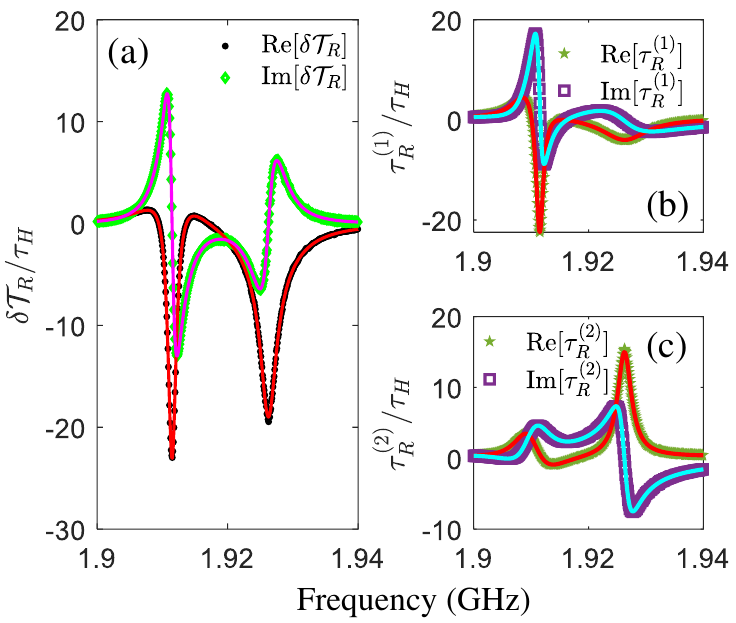}
\caption{Fitting example of reflection time difference/delay for a single pair of shape and Feshbach resonances for a ring graph with $L_1\neq L_2$. (a) shows an example of fitting complex reflection time difference ($\delta \mathcal{T}_R = \tau_R^{(1)} - \tau_R^{(2)}$) experiment data for mode $n=7$. The left feature is due to the Feshbach resonance, while the right one is due to the shape resonance. Parts (b) and (c) demonstrate the reconstruction of the individual reflection time delays on both ports, compared to the data, using the fitted reflection zeros and Wigner poles (see Fig. \ref{Feshbach_Wigner}) information. All time delays are presented normalized by the Heisenberg time $\tau_H$ of the loop graph.}
\label{Reflection}
\end{figure}

We note that Eq. (\ref{FeshPole}) predicts that the resonance width $\Gamma_n$ (imaginary part of the pole) increases as $(c/2\pi)[(2\pi n\delta)^2/(8{\Sigma^3})]$.  Putting the measured values of $\Sigma$ and $\delta$ into this expression gives the red solid curve in Fig. \ref{Feshbach_Wigner}, which demonstrates very good agreement between the data and the prediction in Eq. (\ref{FeshPole}).  Figure \ref{Feshbach_Wigner} also shows the uniform absorption strength $\eta$ increases with frequency.  A more detailed discussion of uniform loss, with comparisons to independent measurements and modeling, can be found in Appendix \ref{appendix:AttenCoax}.  

There is an interesting competition between $\Gamma_n$ and $\eta$ with regards to the complex Wigner time delay in this graph.  Figure \ref{Feshbach_Wigner} shows that $\Gamma_n$ crosses over the value of $\eta$ at approximately mode 27.  Equation (\ref{rWTD}) shows that this will give rise to a change in sign of the nearly-resonant contribution to $\text{Re}[\tau_W]$.  This crossover-related sign change is clearly evident in the full plot of $\text{Re}[\tau_W]$ vs. frequency in Fig. \ref{WignerTimeDelay}.  Further, Fig. \ref{Feshbach_Wigner}(a) shows the fitted real parts of the zeros and poles from the complex Wigner time delay, and they both increase in proportion to $n$, as predicted in Eqs. (\ref{ShapePole}) and (\ref{ShapeZero}) \cite{Waltner2013}.  The solid red line in Fig. \ref{Feshbach_Wigner}(a) shows the prediction based on the measured value of $\Sigma$.  

In Fig. \ref{Feshbach_Transmission}, we plot the fitted imaginary location of the poles (in the form of $\Gamma_n + \eta$) from the complex transmission time delay data together with the previously extracted Wigner poles data from Fig. \ref{Feshbach_Wigner}, and they agree very well.  This validates the hypothesis that the two time delays ($\tau_W$ and $\tau_T$) share the same pole information. Fig. \ref{Feshbach_Transmission} also shows the fitted imaginary parts of the zeros (in the form of $\text{Im}\ t_n - \eta$) from the complex transmission time delay for the Feshbach modes, together with the previously extracted uniform attenuation value ($-\eta$) from Fig. \ref{Feshbach_Wigner}, and they match very well. This implies the transmission zeros are purely real (i.e. $\text{Im}\ t_n =0$), and the data is consistent with this interpretation. Further detailed discussion on the transmission zeros can be found in Appendix \ref{appendix:TZeros}.

\begin{figure}[ht]
\includegraphics[width=86mm]{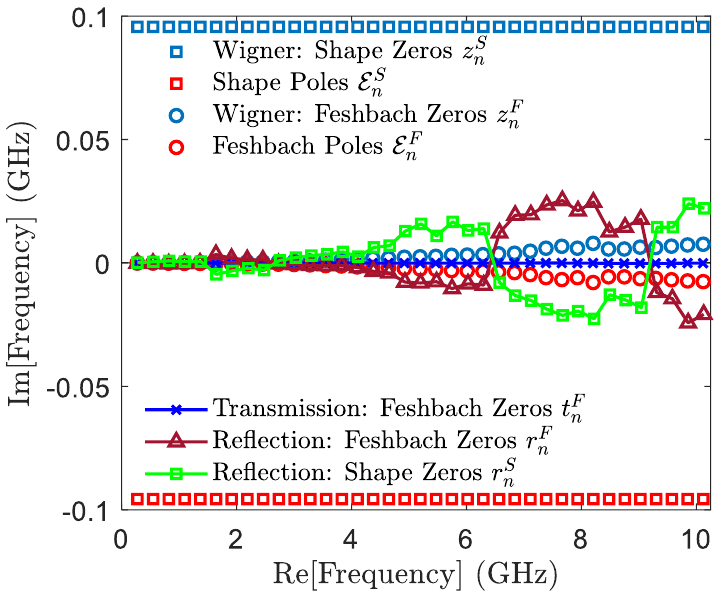}
\caption{Summary of all zeros and poles in the complex frequency plane for shape and Feshbach resonances extracted from Wigner/Transmission/Reflection time delay analysis for the first 37 modes of the microwave ring graph. The Wigner zeros $z_n^S$ (blue squares) and poles $\mathcal{E}_n^S$ (red squares) of the shape resonances are located far from the real axis. The Wigner zeros $z_n^F$ (blue circles) and poles $\mathcal{E}_n^F$ (red circles) of the Feshbach resonances are close to, and symmetrically arrayed about, the real axis. The transmission zeros $t_n^F$ (blue crosses) of the Feshbach resonances lie on the real axis. The reflection zeros $r_n^F \& r_n^S$ of the Feshbach resonances (dark red triangles) and the shape resonances (green squares) are symmetrically arrayed about the real axis.}
\label{ZerosPoles}
\end{figure}

For the reflection time delay analysis, there are two sets of zeros and poles, one each from the shape and Feshbach resonances. One can use the reflection time difference quantity to simplify the analysis, as it contains only the contribution from the zeros. Figure \ref{Reflection} illustrates the reflection time delay/difference analysis process. Figure \ref{Reflection}(a) is an example of fitting the complex reflection time difference to Eqs. (\ref{rRTDD}) and (\ref{iRTDD}) for a single pair of shape and Feshbach resonances. The fitting process was repeated for all $37\times2$ modes utilizing two sets of the reflection zeros ($r_n^F=u_n^F+iv_n^F$ \& $r_n^S=u_n^S+iv_n^S$) as fitting parameters (along with a single value for $\eta$ for each pair), and all fits were very successful. We then examined the complex reflection time delay data for the individual channels, by putting the extracted two sets of reflection zeros ($r_n^F$ \& $r_n^S$) and the previously extracted Wigner poles ($\mathcal{E}_n^F$ \& $\mathcal{E}_n^S$) into the modelling formula Eqs. (\ref{rRTD1}) -- (\ref{iRTD2}). The modelling prediction (with no further fitting adjustments) are plotted with the experimental data in Figs. \ref{Reflection}(b) and \ref{Reflection}(c), and they agree remarkably well.  This indicates that the individual reflection time delays also share the same pole information with the other time delays.

Finally, we present a summary of all zeros and poles extracted from the time delays analysis for the first 37 modes of the microwave ring graph in Fig. \ref{ZerosPoles}.

\section{$S$-matrix reconstruction over the complex plane}
\label{sec:detSAnalysis}

\begin{figure}[ht]
\includegraphics[width=86mm]{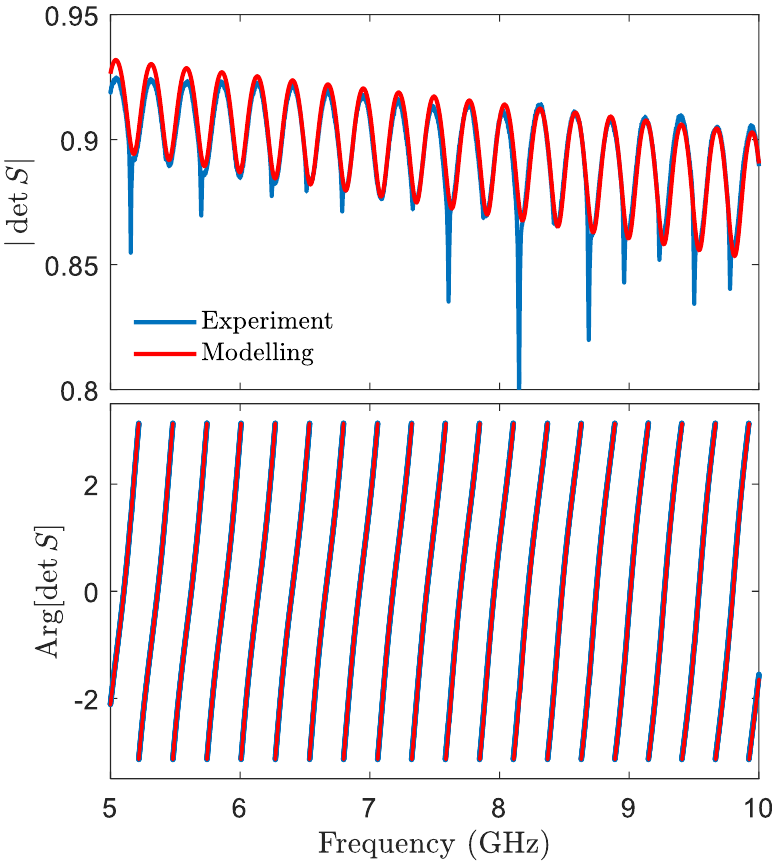}
\caption{Comparison of modelling (red line) and experimental data (blue line) for $\det S$ with shape resonances only in a symmetrical ($L_1=L_2$) ring graph. The modelling data is calculated from Eq. (\ref{detSmodel}) using the Wigner zeros and poles for the shape resonances (see the blue and red squares in Fig. \ref{ZerosPoles}). Upper plot shows the magnitude of $\det S$, while the lower plot shows the phase of $\det S$.}
\label{detS_Shape}
\end{figure}

\begin{figure}[ht]
\includegraphics[width=86mm]{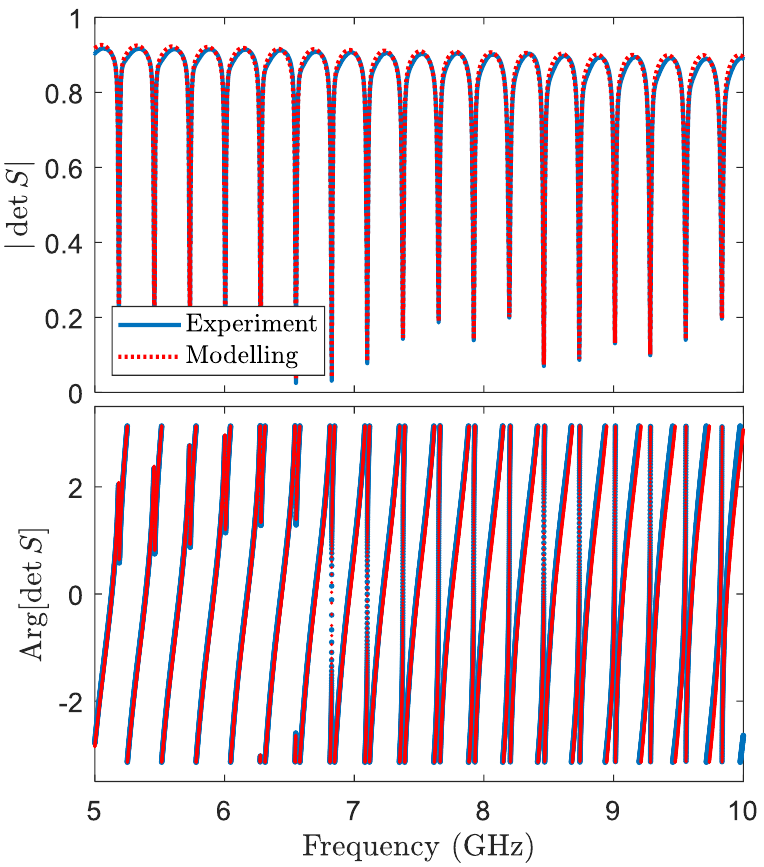}
\caption{Comparison of modelling (red dashed line) and experimental data (blue line) for $\det S$ with both shape and Feshbach resonances in an asymmetrical ($L_1\neq L_2$) ring graph. The modelling data is calculated from Eq. (\ref{detSmodel}) using the Wigner zeros and poles for the shape resonances (see the blue and red squares in Fig. \ref{ZerosPoles}) and the Wigner zeros and poles for the Feshbach resonances (see the blue and red circles in Fig. \ref{ZerosPoles}). Upper plot shows the magnitude of $\det S$, while the lower plot shows the phase of $\det S$.}
\label{detS_Feshbach}
\end{figure}

Now that we have all the zeros and poles information for the scattering system, we would like to examine the modelling for $\det S$ on the real frequency axis utilizing Eq. (\ref{detSmodel}). We reconstructed $\det S$ based on Eq. (\ref{detSmodel}) and the extracted Wigner zeros and poles information summarized in Fig. \ref{ZerosPoles}. Figure \ref{detS_Shape} shows the comparison between the modelling of $\det S$ and the experimental data for a symmetric graph that has the shape resonances only, while Fig. \ref{detS_Feshbach} shows a similar plot with both the Shape and Feshbach resonances present in the scattering system. The modelling agrees very well with the experiment for both the magnitude and phase of $\det S$.  Note that a small delay (0.08 ns) had to be added to the model to show detailed agreement with the data.  We attribute this to about 2.4 cm of un-calibrated transmission line outside of the loop graph, occurring in the third port of each of the tee junctions.

\begin{figure*}[ht]
\includegraphics[width=\textwidth]{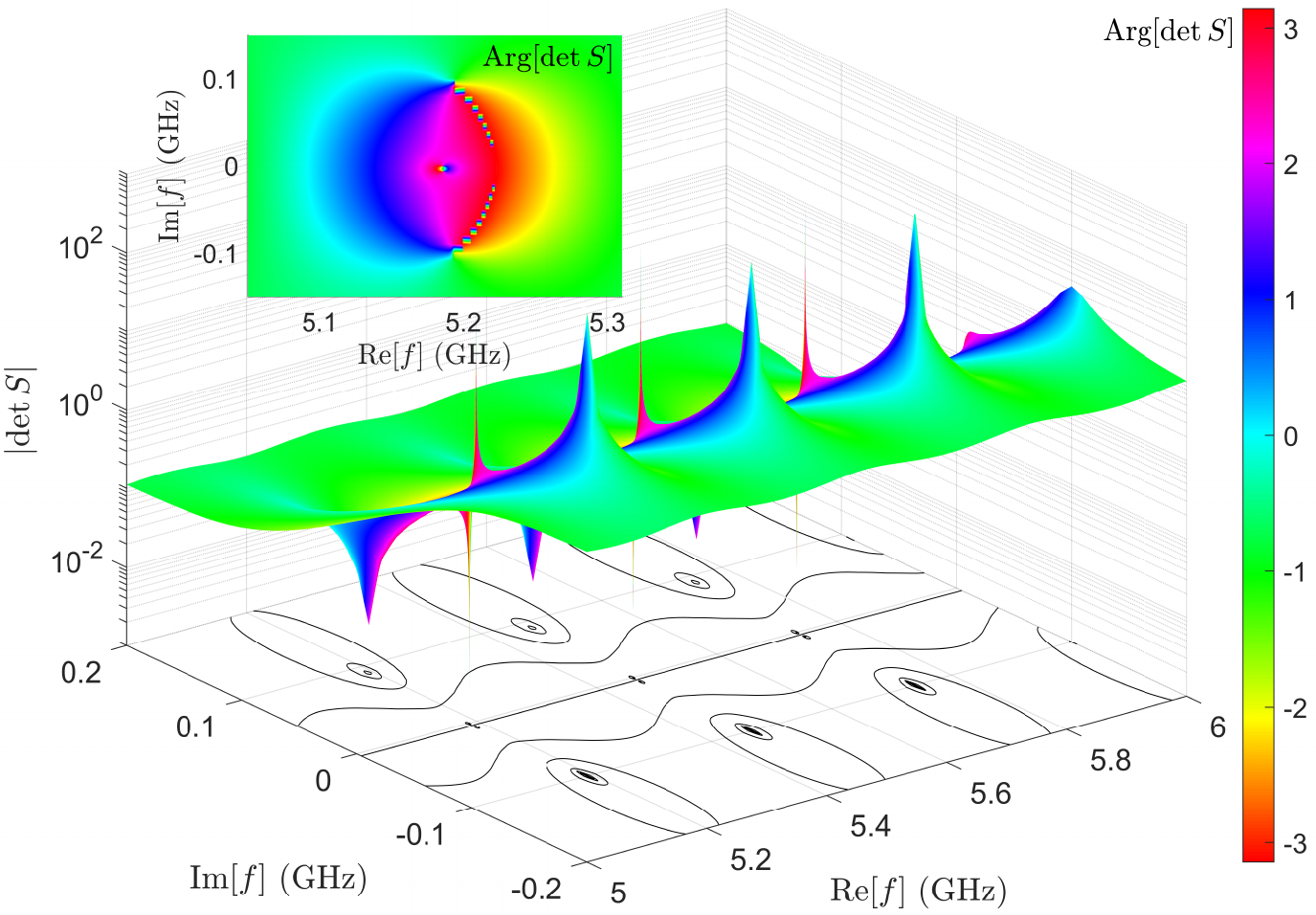}
\caption{Complex representation of $\det S$ evaluated over the complex frequency plane for several modes of an asymmetric ($L_1 \neq L_2$) ring graph.  $\det S$ is calculated from Eq. (\ref{detSmodel}) using complex frequency and the Wigner zeros and poles for the shape resonances (see the blue and red squares in Fig. \ref{ZerosPoles}) and the Wigner zeros and poles for the Feshbach resonances (see the blue and red circles in Fig. \ref{ZerosPoles}). The 3D plot represents $|\det S|$ on a log scale and reveals the zeros (dips) and poles (peaks) at different locations in complex frequency. The base plane shows contour lines of the magnitude of $|\det S|$ in the complex frequency plane.  The colorbar on the right shows the phase of the constructed $\det S$.  The inset shows a 2D top view of $\text{Arg}[\det S]$ for a single pair of shape and Feshbach zeros and poles.}
\label{detS_complex}
\end{figure*}

Reconstructing the $S$-matrix over the entire complex frequency plane is generally difficult to accomplish experimentally.  Here we construct complex $\det S$ on the complex frequency plane ($E$ or $f$ being complex) by continuation of Eq. (\ref{detSmodel}), along with the extracted Wigner zeros and poles information. Fig. \ref{detS_complex} (and Fig. \ref{detS_complex_v2}) shows a 3D reconstruction of the complex $\det S$ for an asymmetric ring graph evaluated over the complex frequency plane with both the shape and Feshbach resonances present. We can see a series of dips and peaks, which reveal the zero and pole locations in the complex frequency domain.

\begin{figure}[ht]
\includegraphics[width=86mm]{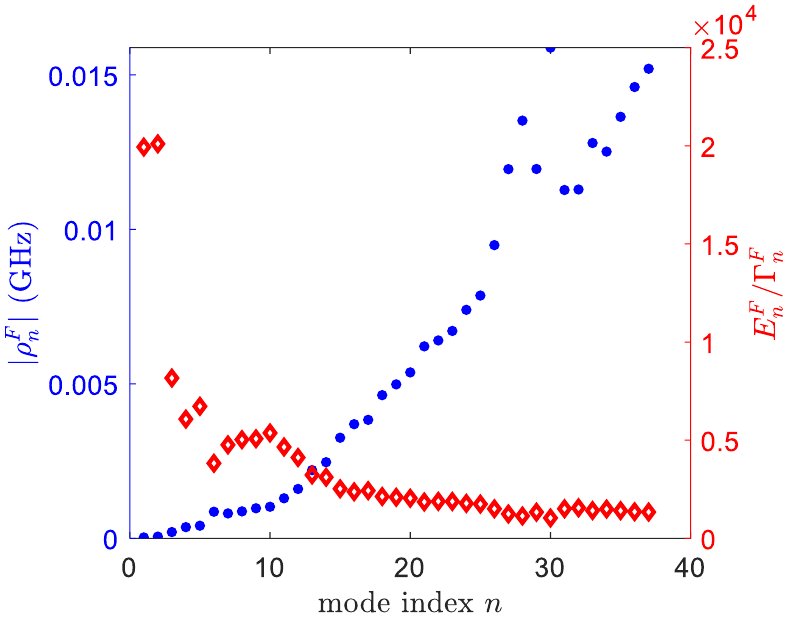}
\caption{Plot of residue $\rho_n^F$ and the `quality factor' of the Feshbach poles, versus mode index $n$, for an asymmetric microwave loop graph.  The blue filled circles show the absolute magnitude of the residue $|\rho_n^F|$ as a function of mode index based on the extracted Feshbach poles and zeros, while the red open diamonds show the associated ratio of $E_n^F/\Gamma_n^F$ of the Feshbach poles.}
\label{Residue}
\end{figure}

Other methods exist for $S$-matrix reconstruction.  One approach is to use harmonic inversion, in which frequency domain data is transformed into the time domain and fit to a time-decay made up of a sum of many poles \cite{Mand97,Wier08,Kuhl08}.  This technique is quite successful for finding poles, but does not directly determine the zeros of the $S$-matrix.  Note that complex time delay can be used to augment a harmonic inversion search for $S$-matrix poles.  Another approach to finding scattering poles is to use numerical methods to find outgoing-only solutions to wave equations in terms of quasinormal modes, and therefore identify the complex pole positions \cite{Ching98,Kris20}.  A more complete approach is to use Weierstrass factorization of the $S$-matrix, and to also include solutions to the wave equations that involve ingoing-only solutions to identify the zeros of $S$ \cite{Grig13,Grig13b}.  This approach allows one to re-expresses the scattering matrix in terms of a sum of Lorentzians due to the poles, with residues that depend on both the zeros and the poles.  Note that here we retrieve only $\det[S]$, but the full $S$ matrix can also be reconstructed \cite{Grig13,Grig13b}.\\

If a passive zero loss system hosts an embedded eigenstate, i.e., a mode with zero-decay rate, the corresponding $S$-matrix pole will lie on the real frequency axis. In a passive system with finite loss, this is only possible if there is also a degenerate $S$-matrix zero occurring at the same real frequency, where they merge and cancel each other \cite{Mont14,Kras19,Sako20}.  This seems to describe the Feshbach poles and zeros of the ring graph in the limit as $n \rightarrow 0$.  To measure the degree of coincidence of the pole and zero, we can evaluate the residue of the Feshbach poles as a function of mode number.  The residue of $\det[S]$ due to a single (assumed simple) Feshbach pole is given by $\rho_n^F = \det[S(\mathcal{E})] (\mathcal{E} - \mathcal{E}_n^{F, asymm})|_{\mathcal{E} \rightarrow \mathcal{E}_n^{F, asymm}}$.  This in turn can be written as $\rho_n^F \propto \frac{\mathcal{E} - z_n^{F, asymm}}{\mathcal{E} - \mathcal{E}_n^{F, asymm}} (\mathcal{E} - \mathcal{E}_n^{F, asymm})|_{\mathcal{E} \rightarrow \mathcal{E}_n^{F, asymm}} = \mathcal{E}_n^{F, asymm} - z_n^{F, asymm}$, which is just the distance between the Feshbach pole and zero.  Figure \ref{Residue} shows the absolute magnitude of $\rho_n^F$ as a function of mode number based on the extracted Feshbach poles and zeros.  It is clear that in the limit of index going to zero that the pole and zero approach each other, consistent with the development of an embedded eigenstate.  Also shown in Fig. \ref{Residue} is the associated `Q' value of the pole in terms of the ratio $E_n^F/\Gamma_n^F$ of the modes.

\section{Discussion}
\label{sec:Discuss}



Our comprehensive discussion of Wigner, transmission, and the reflection complex time delays in section \ref{sec:compTDs} of the paper gives us the opportunity to address the question: what is the general strategy to maximize the real part of all the complex time delays?  From Eq. (\ref{rWTD}) we see that the real part of $\tau_W$ is maximized when the imaginary part of a scattering pole $\Gamma_n$ is equal to the uniform attenuation rate $\eta$.  This divergence of the Wigner time delay has been previously demonstrated in the context of coherent perfect absorption by several groups \cite{Chen2021gen,Hougne20}.  Also, for the microwave ring graph studied here, we see from the plot of $\tau_W$ vs. frequency in Fig. \ref{WignerTimeDelay} that this condition is nearly met somewhere around 7 GHz.  With tuning of either $\delta$ and/or $\eta$ we could achieve the divergence of $\text{Re}[\tau_W]$ for one or more modes.  

From Eq. (\ref{rTTD}) we see that the real part of $\tau_T$ is maximized when the imaginary part of a transmission zero $\text{Im}[t_n]$ is equal to the uniform attenuation rate $\eta$.  In our data on the microwave ring graph, the imaginary part of the transmission zero is always negative and much smaller in magnitude than the uniform attenuation, so the associated divergence is not visible here.  The data for complex $\tau_T$ vs. frequency for all 37 modes is shown in Fig. \ref{TransmissionTimeDelay}.  The transmission time delay shows nearly sinusoidal oscillations arising from the shape modes, and a series of spikes arising from the Feshbach modes.  As expected, the transmission time delays are generally small in magnitude and show no irregular variations associated with a near degeneracy of $\text{Im}[t_n]$ and $\eta$.   

Finally, from Eqs. (\ref{rRTD1}), (\ref{rRTD2}), and (\ref{rRTDD}) we see that the real part of either $\tau_R^{(1)}$ or $\tau_R^{(2)}$, and the magnitude of $\delta \mathcal{T}_R = \tau_R^{(1)}-\tau_R^{(2)}$, is maximized when the imaginary part of a reflection zero $v_n$ is equal to either plus or minus the uniform attenuation rate, $\pm \eta$.   For our microwave ring graph, we see from the plots of complex $\tau_R$ vs. frequency in Fig. \ref{ReflectionTimeDelay} that this condition is nearly met for a number of modes, including modes 1 and 14.  The extreme values of reflection time delay, on the order of hundreds of Heisenberg times, dwarfs those of the Wigner and transmission times.  In this case we have $v_1^F = -8.65 \times 10^{-5}$ GHz, $v_1^S = 1.05 \times 10^{-4}$ GHz and $\eta = 3.79 \times 10^{-5}$ GHz for mode 1, and $v_{14}^F = 0.0010$ GHz, $v_{14}^S = 0.0045$ GHz and $\eta = 0.0044$ GHz for mode 14, resulting in large values for the real and imaginary parts of $\tau_R$. 

To summarize, we note that divergences in all time delays can be tuned into existence through variation of uniform attenuation $\eta$, or perturbations that systematically vary $E_n$, $\Gamma_n$, $t_n$, or $r_n$. 

What is the practical limit for the maximum value of time delay?  Constructing time delay from experimental $S$-parameter data requires two nearby data points with which we calculate a finite difference approximation to the derivative of $\ln (\det [S])$.  However, the singularity is at a single point in frequency, hence we can never achieve the true divergence this way, although we can get arbitrarily close by taking finer steps in parameter space.  On the other hand, one can tune to the CPA condition of a physical system containing a non-zero loss and create an unbounded time delay at one frequency, as demonstrated with CPA experiments in microwave graphs \cite{Chen2021gen}. 

The introduction of complex time delay analysis now offers the opportunity to study the detailed evolution of poles and zeros in the complex plane when scattering systems are subjected to a variety of perturbations.  A number of methods to controllably drive poles and zeros around the complex plane have been developed in different contexts. As an example in the case of the ring graph, several authors have examined the question of what trajectory an embedded eigenvalue pole leaves the real axis as the ring graph is perturbed \cite{Exner10,Lee16,Lawn21}.  Another opportunity is the manipulation of reflection zeros in the complex frequency plane for multi-port scattering systems to create what are known as reflectionless scattering modes (RSM) \cite{Dhia18,Swee20}.  Reflection ($\tau_R$) and reflection difference ($\delta \mathcal{T}_R$) complex time delays will enable monitoring of reflection zeros so that they can be tuned to the real axis to establish RSMs.\\


Wave chaotic systems have scattering properties that are very sensitive to changes in boundary conditions.  This makes such systems well suited to act as sensors of perturbation, such as motion or displacement of objects located in the scattering domain, through the concept of scattering fidelity \cite{Sch05a,Sch05b,Bini09,Bini10,Bini13}.  In addition, there exists a class of sensors that are based on the coalescence of two or more eigenmodes \cite{Wier14,Hod17}.  In all cases, the longer the dwell time of a wave in a monitored space, the greater its sensitivity to small perturbations \cite{Hougne20,Hougne21}. 

Finally, we discuss a number of important issues associated with our approach to modeling the complex time delays.  In this paper we have taken two distinctly different approaches to modeling the measured time delay.  In the case of the shape resonances, the poles and zeros are relatively far removed from the real axis; the ratio of imaginary part of the pole (and zero) to the mean spacing is approximately $\Gamma_n^S/\Delta E_n^S \sim 0.35$.  In this case, many poles and zeros contribute to the Wigner time delay (as an example) at any given point on the real frequency axis.  For this reason, we fit all of the pole and zero locations at once for the data in Fig. \ref{Shape_WignerTransmission}.  In addition, the product over modes in Eq. (\ref{detSmodel}) extends over $\pm 200$ modes in order to properly reproduce $\det S$ in Figs. \ref{detS_Shape} and \ref{detS_Feshbach}.  On the other hand, when poles and zeros are close to the real axis, it is possible to treat each pole/zero pair individually.  This is the case for the Feshbach resonances where we find the ratio of imaginary part of the pole to the mean spacing is roughly $\Gamma_n^F/\Delta E_n^F \sim 0.01$.  In this case the contribution to the time delay in a given narrow frequency window is dominated by the nearest pole and zero.  This is the case for the fits shown in the insets of Figs. \ref{Feshbach_Wigner} and \ref{Feshbach_Transmission}, and the fits shown in Fig. \ref{Reflection}.  We have checked this assumption by a number of methods.  First, our correct recovery of the measured $\det S$ on the real axis, as shown in Fig. \ref{detS_Feshbach}, is a clear test of the assumption that the fitting of individual Feshbach poles and zeros is adequate to model the global scattering matrix at arbitrary real frequencies.  Secondly, we have checked that adding terms to the complex time delay arising from neighboring poles and zeros has no effect on our fitting of individual mode data, such as those shown in the insets of Figs. \ref{Feshbach_Wigner} and \ref{Feshbach_Transmission}.

There is one additional potential limitation of the above description of complex time delay.  Assuming a single uniform value of the loss parameter $\eta$ at a given frequency is an approximation, especially for our ring graph. The graph has a variable phase shifter in it that is not a homogeneous transmission line.  There may be point-like loss centers that exist in this microwave graph, which we are not modelling properly with just a uniform attenuation.  Also, in the fitting of complex time delay vs. frequency, we assume that the value of $\eta$ is constant in the narrow frequency range around each pair of shape/Feshbach modes (as in Fig. \ref{Reflection}), although we believe that this is a good approximation for the data and analysis presented here. 

\section{Conclusions}
\label{sec:Conclusions}

We provide a comprehensive analysis of the ring graph scattering response in terms of poles and zeros of the $S$-matrix, and the reflection and transmission submatrices.  We have treated the complex Wigner-Smith, reflection and transmission time delays on equal footing, all in one experimental setting.  We also create a faithful reconstruction of the complex determinant of the $S$-matrix over the complex frequency plane from the experimentally extracted poles and zeros.  More generally, we provide the first comprehensive treatment of complex Wigner, transmission, reflection, and reflection difference time delays.  We also provide a prescription for maximizing the real part of all complex time delays in terms of the poles and zeros of the scattering matrix, and the uniform attenuation in the system. \\

Acknowledgements: We gratefully acknowledge discussions with Yan V. Fyodorov and Uzy Smilansky.  This work was supported by ONR Grant No. N000141912481, and DARPA WARDEN Grant No. HR00112120021.

\appendix

\section{Additional Data}
\label{appendix:AddiData}

\begin{figure*}[ht]
\includegraphics[width=\textwidth]{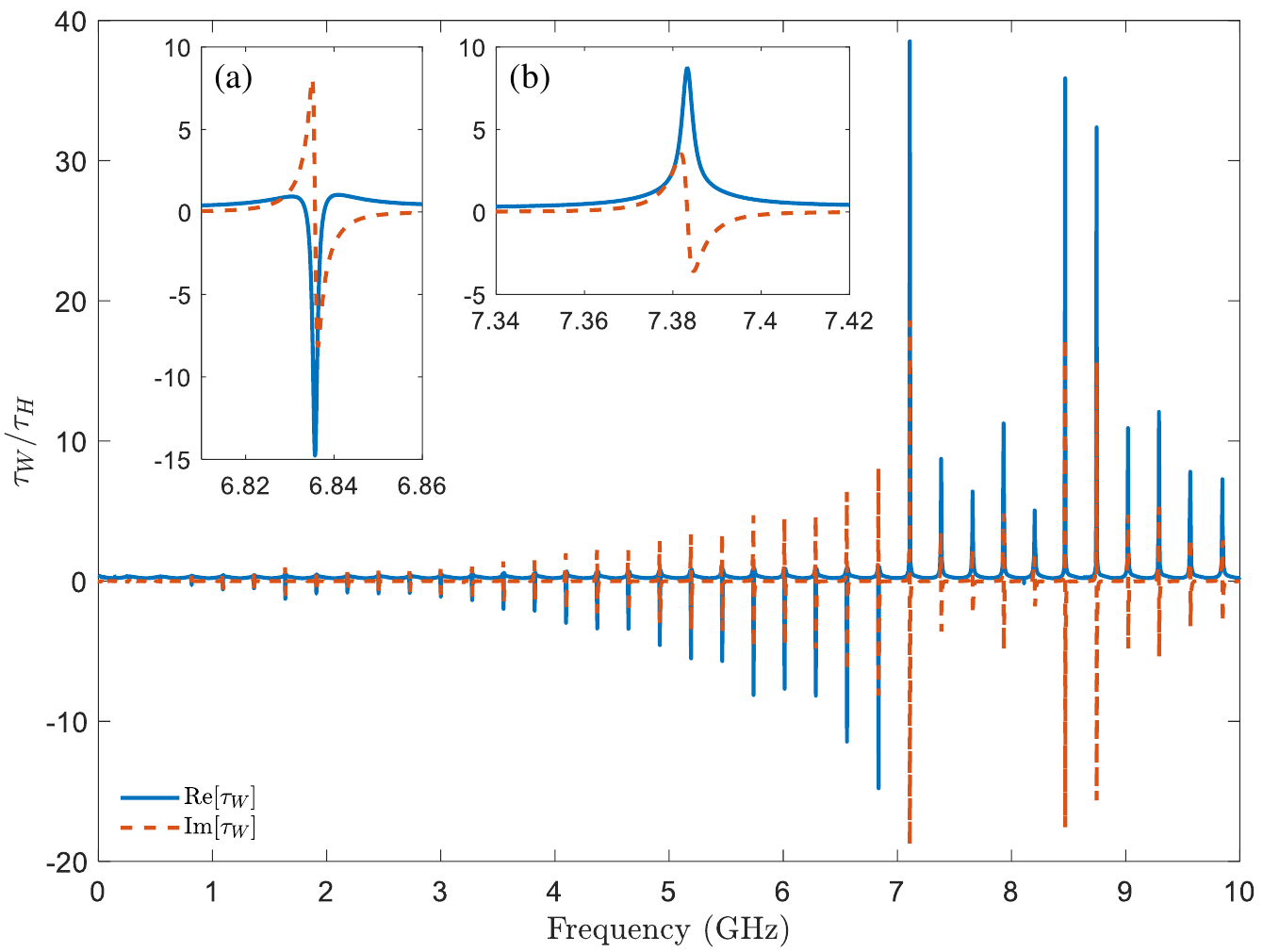}
\caption{Complex Wigner time delay $\tau_W$ (normalized by the Heisenberg time $\tau_H$) determined from measured $S$-matrix data for 37 modes ($0-10$ GHz) in an asymmetrical ($L_1\neq L_2$) microwave ring graph.  The extreme values of $\tau_W$ are dominated by Feshbach resonances.  Note the sign change of the $\text{Re}[\tau_W]$ extreme values near 7 GHz, which corresponds to the crossover between $\Gamma_n$ and $\eta$ in Fig. \ref{Feshbach_Wigner}. Insets (a) and (b) show zoom-in details of the complex Wigner time delay for individual modes on either side of the crossover.}
\label{WignerTimeDelay}
\end{figure*}

\begin{figure*}[ht]
\includegraphics[width=\textwidth]{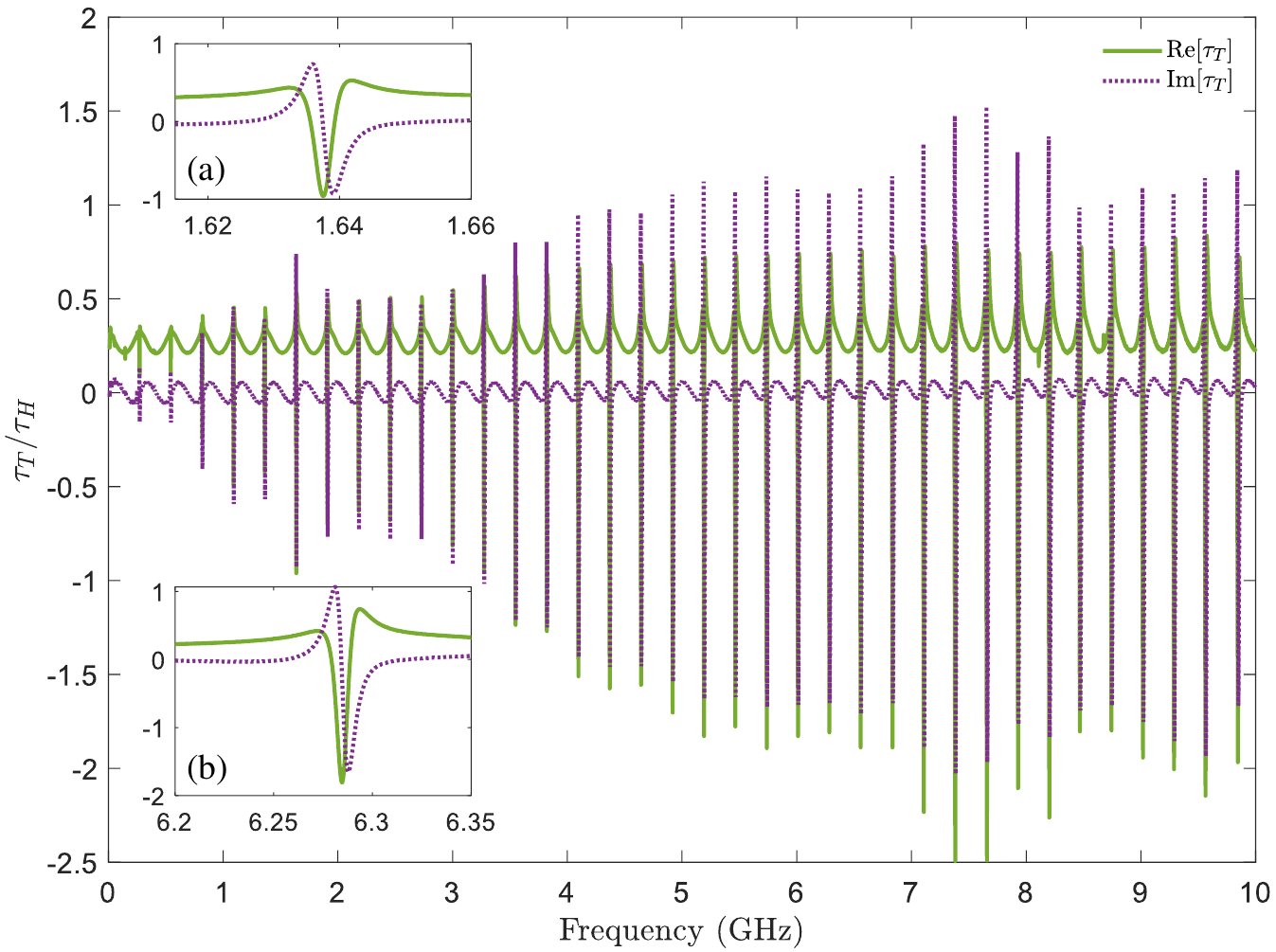}
\caption{Complex transmission time delay $\tau_T$ determined from measured $S$-matrix data for 37 modes ($0-10$ GHz) in an asymmetrical ($L_1\neq L_2$) microwave ring graph normalized by the Heisenberg time $\tau_H$.  The extreme values of $\tau_T$ are dominated by Feshbach resonances.  The nearly sinusoidal variations of $\text{Re}[\tau_T]$ and $\text{Im}[\tau_T]$ with frequency are due to the shape resonances.  Insets (a) and (b) show the zoom-in details of the complex transmission time delay for two individual modes.}
\label{TransmissionTimeDelay}
\end{figure*}

\begin{figure*}[ht]
\includegraphics[width=\textwidth]{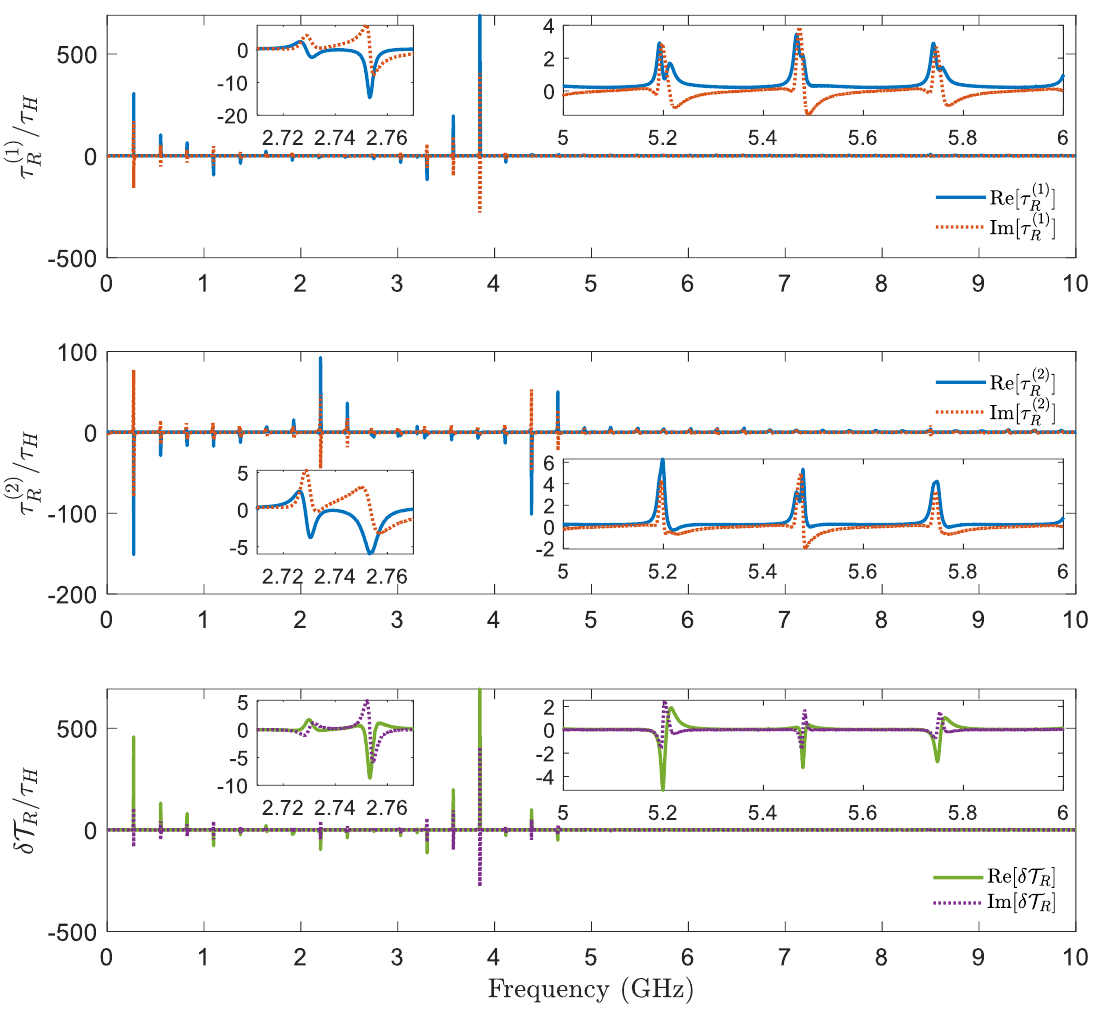}
\caption{Complex reflection time delays $\tau_R^{(1)}$, $\tau_R^{(2)}$ and their difference $\delta \mathcal{T}_R = \tau_R^{(1)} - \tau_R^{(2)}$ determined from measured $S$-matrix data for 37 modes ($0-10$ GHz) in an asymmetrical ($L_1\neq L_2$) microwave ring graph, normalized by the Heisenberg time $\tau_H$.  Insets show the zoom-in details of the complex reflection time delay/difference for individual sets of shape and Feshbach modes.}
\label{ReflectionTimeDelay}
\end{figure*}

Here we present the complex Wigner-Smith ($\tau_W$) (Fig. \ref{WignerTimeDelay}), transmission ($\tau_T$) (Fig. \ref{TransmissionTimeDelay}), and reflection ($\tau_R$) (Fig. \ref{ReflectionTimeDelay}) time delays over the full measurement frequency range (0 -- 10 GHz), including all 37 modes of the asymmetrical ($L_1\neq L_2$) microwave ring graph.  Examining the complex time delays over a broad range of frequency brings out new aspects of the data, as discussed in Section \ref{sec:Discuss}.

Figure \ref{WignerTimeDelay} shows the complex Wigner time delay extracted from the experiment over the entire measurement frequency range.  We have already noted in Section \ref{sec:DataAnalysis} the change in sign of $\text{Re}[\tau_W]$ as a function of frequency due to the crossover of the imaginary part of the Feshbach pole $\Gamma_n$ and the uniform attenuation $\eta$.  Another feature to note is that the shape resonances produce a relatively small variation in $\tau_W$ compared to the sharp features arising from the Feshbach modes.  Both features together create time delays on the scale of at most 10's of Heisenberg times in this particular experimental realization and frequency range.

Figure \ref{TransmissionTimeDelay} shows the complex transmission time delay extracted from the experiment over the entire measurement frequency range.  We note that the magnitude of the transmission time delays are limited in magnitude to approximately 2 times the Heisenberg time in this case.  The reason for such small variations is that the transmission time delays have contributions from both the zeros and the poles, and the two contributions have similar magnitudes but opposite signs.  Thus the resulting transmission time delays are rather small compared to $\tau_W$ and $\tau_R$. Further detailed discussion of $\tau_T$ is given in Appendix \ref{appendix:TZeros}.

The reflection time delays shown in Fig. \ref{ReflectionTimeDelay} show significantly larger range of variation as compared to the Wigner and transmission time delays. To see why this is the case, we can examine Eqs. (21) -- (24), which model the behavior of the reflection time delays.  One can see that the width and the extreme value of the first Lorentzian term is determined by $|v_n \pm \eta|$. The reflection zeros $r_n=u_n+iv_n$ are the complex eigenvalues of $H+i(\Gamma_{W}^{(1)} - \Gamma_{W}^{(2)})$. In our experimental setup, we have very similar coupling properties for ports 1 and 2, i.e. $\Gamma_{W}^{(1)} \approx \Gamma_{W}^{(2)}$. Thus, the imaginary part of the reflection zeros $v_n$ should be fairly small. At low frequencies, the uniform attenuation $\eta$ is also very small, and is comparable to $v_n$. This leads to a very small width of the Lorentzian resonance, which in turn produces very large extreme values of the reflection time delay, on the order of 100's of Heisenberg times, at low frequencies. At larger frequencies, however, the uniform attenuation $\eta$ becomes fairly large, and dominates the width of the Lorentzian resonance. Therefore, the reflection time delays change back to the order of a few Heisenberg times.

\section{Uniform Attenuation Estimation for Coaxial Cable}
\label{appendix:AttenCoax}

We estimate the uniform attenuation $\eta$ in the ring graph system both theoretically and experimentally. From \cite{pozar2011microwave}, we derived the corresponding expression for the uniform attenuation ($\Gamma$) of a homogeneous coaxial cable, expressed in terms of an angular frequency: 
\begin{align}
    \label{etaModel}
    \Gamma = \frac{1}{2}\left[2\pi f\tan \delta + \sqrt{\frac{2\pi f\rho}{2\mu_0}} \frac{1}{\sqrt{\epsilon_r}} \frac{1}{\ln{(b/a)}} (\frac{1}{a} + \frac{1}{b}) \right],
\end{align}
where $f$ is the linear frequency, $\tan \delta = 0.00028$ and $\epsilon_r = 2.1$ are the dielectric loss tangent the relative dielectric constant of the Teflon dielectric, $\rho = 4.4 \times 10^{-8} \ \Omega \cdot m$ is the resistivity of the metals in the cable, $\mu_0 = 4\pi \times 10^{-7}$ H/m is the permeability of vacuum, and $a = 0.46 \times 10^{-3}$ m and $b = 1.49 \times 10^{-3}$ m are the radii of the inner and outer conductors, respectively.  These values are typical for the coaxial cables used in our experiments.

\begin{figure}[ht]
\includegraphics[width=86mm]{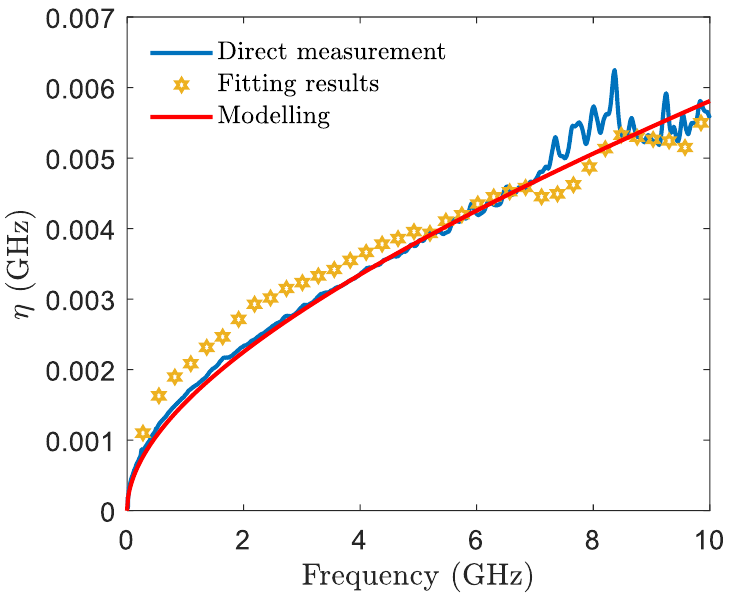}
\caption{Comparison of three different ways to determine the uniform attenuation of the loop graph: by means of direct measurement of insertion loss through $S_{21}$, fitting results to complex time delays ($\eta$), and direct modelling ($\Gamma$).  The blue line shows the data obtained by measuring the $S_{21}$ insertion loss of a serial connection of the coaxial cable and the phase shifter shown in Fig. \ref{schematic}(b).  The yellow stars show the fitting results for $\eta$ from the complex Wigner time delay analysis in Fig. \ref{Feshbach_Wigner}.  The red line shows the theoretical modelling (Eq. (\ref{etaModel})) of $\Gamma/2\pi$ in a coaxial cable.}
\label{eta_estimation}
\end{figure}

We also performed a direct measurement of the uniform attenuation for the components making up the ring graph.  We connected the coaxial cable and the phase shifter from Fig. \ref{schematic}(b) in series and measured the transmission $S_{21}$ insertion loss as a function of frequency.  The comparison of uniform attenuation between direct measurement (from $S_{21}$), fitting results ($\eta$) and the modelling ($\Gamma$) is plotted in Fig. \ref{eta_estimation}.  The agreement between these three independent estimates is reasonably good.  Note that the coaxial phase shifter is not a uniform coaxial structure, and evidence of internal resonances are visible in Fig. \ref{eta_estimation} above 7 GHz.   Note that the fit $\eta$ values are slightly higher than the direct loss measurement below 7 GHz, but then are slightly lower above that frequency.  This comparison gives us confidence that the values of $\eta$ extracted from complex time delay analysis are quite reasonable.

\section{Transmission Zeros}
\label{appendix:TZeros}

In the transmission zeros analysis for the Feshbach resonances, we fit the experimental data to Eqs. (\ref{rTTD}) and (\ref{iTTD}), after removing the contributions from the shape resonances. We may rewrite the complex transmission time delay as $\tau_T = \tau_T^Z + \tau_T^P$ \cite{Genack21}, where $\tau_T^Z$ and $\tau_T^P$ are the contributions from zeros and poles, respectively. Then Eqs. (\ref{rTTD}) and (\ref{iTTD}) can be rewritten as
\begin{align}
    \label{rTTD_Z}
    \text{Re}\ \tau_T^Z(E;\eta) &= \sum_{n=1}^{N-M} \frac{\text{Im}\ t_n - \eta}{(E-\text{Re}\ t_n)^2 + (\text{Im}\ t_n - \eta)^2}, \\
    \label{iTTD_Z}
    \text{Im}\ \tau_T^Z(E;\eta) &= - \sum_{n=1}^{N-M} \frac{E-\text{Re}\ t_n}{(E-\text{Re}\ t_n)^2 + (\text{Im}\ t_n - \eta)^2}, \\
    \label{rTTD_P}
    \text{Re}\ \tau_T^P(E;\eta) &= \sum_{n=1}^{N} \frac{\Gamma_n + \eta}{(E-E_n)^2 + (\Gamma_n + \eta)^2}, \\
    \label{iTTD_P}
    \text{Im}\ \tau_T^P(E;\eta) &= \sum_{n=1}^{N} \frac{E - E_n}{(E-E_n)^2 + (\Gamma_n + \eta)^2}.
\end{align}

\begin{figure}[ht]
\includegraphics[width=86mm]{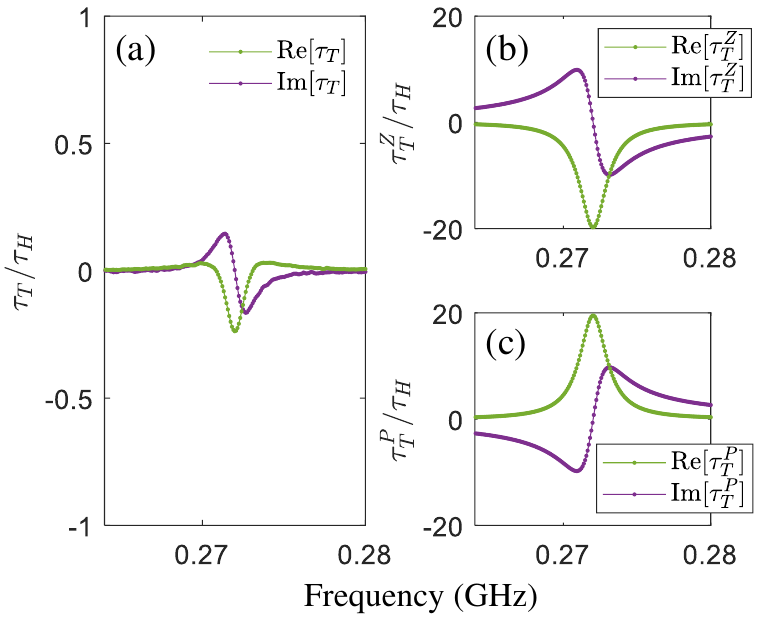}
\caption{Complex transmission time delay data for a single Feshbach mode ($n = 1$) and its contributions from zeros and poles. (a) shows the total complex transmission time delay data ($\tau_T$), while (b) and (c) show the contribution from the zero ($\tau_T^Z$) and the pole ($\tau_T^P$), respectively. Here $\tau_T$ is from experimental data, while $\tau_T^P$ is calculated based on Eqs. (\ref{rTTD_P}) \& (\ref{iTTD_P}) with the pole information extracted from the complex Wigner time delay analysis (see Fig. \ref{Feshbach_Wigner}). $\tau_T^Z$ is obtained by $\tau_T^Z = \tau_T - \tau_T^P$.}
\label{tauT_Zero_Pole}
\end{figure}

We plot $\tau_T^Z$ and $\tau_T^P$ for a single Feshbach mode ($n = 1$) in Fig. \ref{tauT_Zero_Pole}.  Here $\tau_T^P$ is calculated using the pole information extracted from the complex Wigner time delay analysis (see Fig. \ref{Feshbach_Wigner}), since all three time delays share the same poles. $\tau_T^Z$ can then be obtained through $\tau_T^Z = \tau_T - \tau_T^P$, where $\tau_T$ is the experimental data. Fig. \ref{tauT_Zero_Pole} shows that $\tau_T^Z$ and $\tau_T^P$ are approximately equal in magnitude, both much larger than $\tau_T$, but have opposite signs. From \cite{Genack21,Genack22} we learned that the transmission zeros $t_n$ will be on the real axis, i.e. $\text{Im}[t_n] = 0$, such that $\text{Im}[t_n] - \eta = -\eta$. For this ($n = 1$) Feshbach mode, the imaginary part of the pole $\Gamma_n$ is very small compared to the uniform attenuation $\eta$ (see Fig. \ref{Feshbach_Wigner}), thus we have $\Gamma_n + \eta \approx \eta$. Under such conditions, Eqs. (\ref{rTTD_Z}) -- (\ref{iTTD_P}) can be written as $\text{Re}[\tau_T^Z]_{n=1} = - \eta / [(E-\text{Re}\ t_n)^2 + \eta^2]$, $\text{Re}[\tau_T^P]_{n=1} \approx + \eta / [(E-E_n)^2 + \eta^2]$, $\text{Im}[\tau_T^Z]_{n=1} = - (E-\text{Re}\ t_n) / [(E-\text{Re}\ t_n)^2 + \eta^2]$, and $\text{Im}[\tau_T^P]_{n=1} \approx (E-E_n) / [(E-E_n)^2 + \eta^2]$. Since $\text{Re}\ t_n \approx E_n$, we then arrive at $[\tau_T^Z]_{n=1} \approx - [\tau_T^P]_{n=1}$, which is consistent with what is shown in Fig. \ref{tauT_Zero_Pole}. This also explains why $\tau_T = \tau_T^Z + \tau_T^P$ is so small for this Feshbach mode ($n = 1$) (see Fig. \ref{tauT_Zero_Pole}(a)).

\begin{figure}[ht]
\includegraphics[width=86mm]{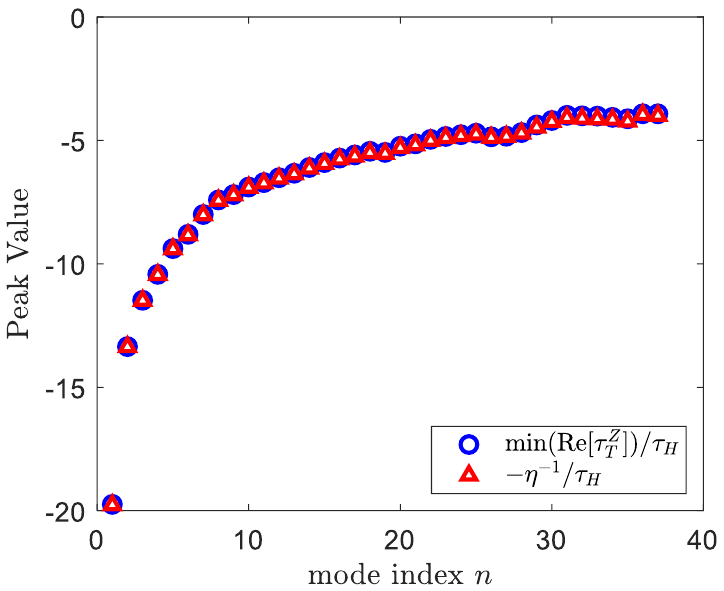}
\caption{Comparison between the peak value of $\text{Re}[\tau_T^Z]$ and $-\eta^{-1}$ for all 37 modes of the microwave ring graph. Blue circles show the peak value of $\text{Re}[\tau_T^Z]$ from experimental data, while red triangles show $-\eta^{-1}$ calculated from the data in Fig. \ref{Feshbach_Wigner}. Both quantities are presented normalized by the Heisenberg time $\tau_H$ of the loop graph.}
\label{tauT_Z_peak}
\end{figure}

When analyzing the transmission time delay data, one may assume either a single zero or a conjugate pair of zeros in the modelling \cite{Genack21,Genack22}. We tried using a conjugate pair of zeros to fit the data, but were unable to achieve reasonable fitting results. A pair of zeros would contribute to the real part of transmission time delay with a local extremum at $E = \text{Re}\ t_n$ of $\text{Re}[\tau_T^Z] = \frac{2\eta}{(\text{Im}\ t_n)^2 - \eta^2}$. Unfortunately this expression demands negative values for $(\text{Im}\ t_n)^2$ for our data, therefore the pair of zeros assumption is inconsistent with the data. On the other hand, the contribution of a \textit{single} zero to $\text{Re}[\tau_T]$ is $\text{Re}[\tau_T^Z] = \frac{-\eta}{(E - \text{Im}\ t_n)^2 - \eta^2}$, with peak value $-\eta^{-1}$. We plot the comparison between the peak value of $\text{Re}[\tau_T^Z]$ (from data) vs $-\eta^{-1}$ (from Fig. \ref{Feshbach_Wigner}) for all 37 modes in Fig. \ref{tauT_Z_peak}, and they agree extremely well, justifying our single-zero hypothesis.  In summary, placing all of the transmission zeros on the real axis is consistent with the data.\\

\section{Additional $\det[S]$ Reconstruction Plot}
\label{appendix:Addi3DPlot}

\begin{figure*}[ht]
\includegraphics[width=\textwidth]{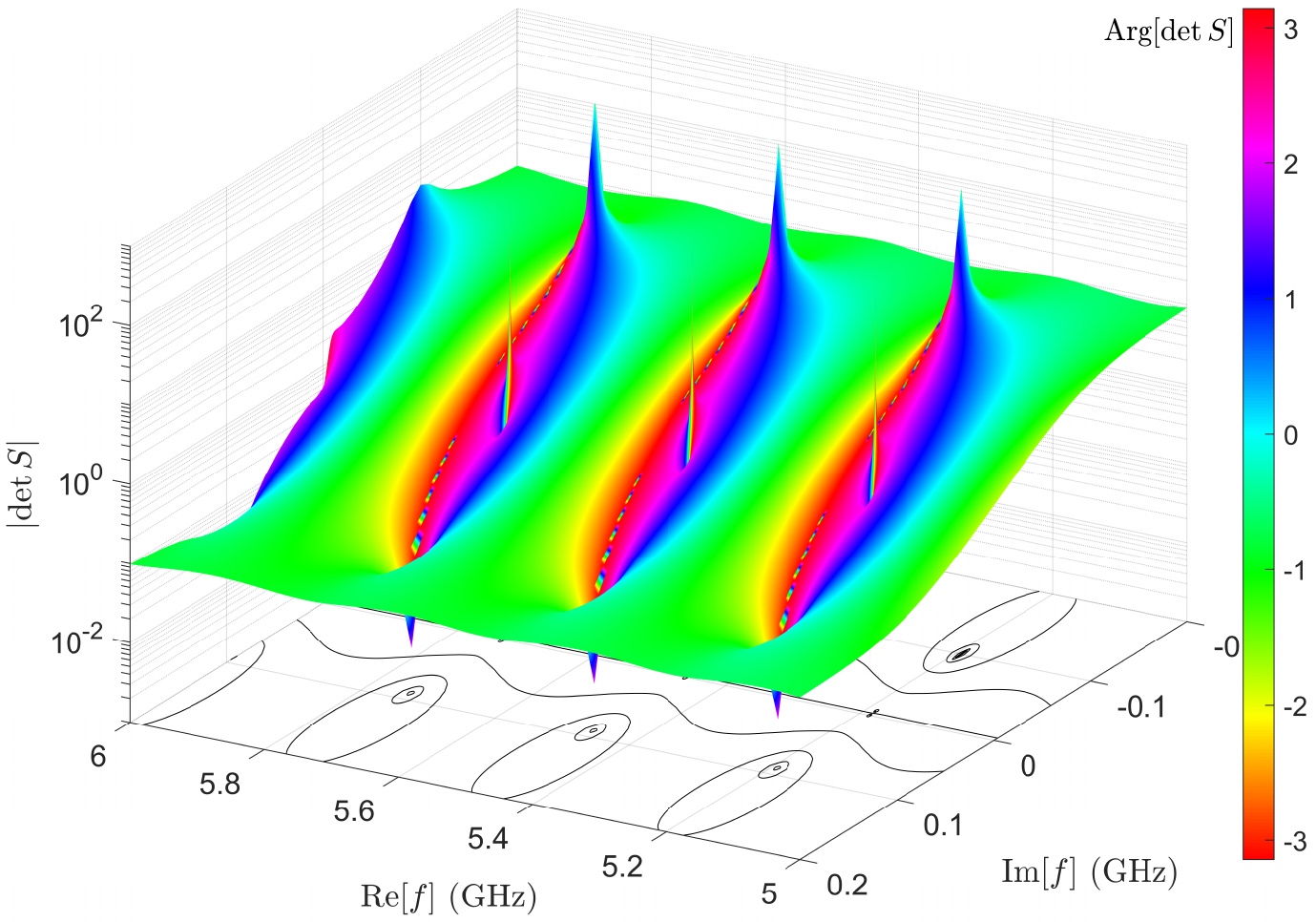}
\caption{Complex representation of $\det S$ evaluated over the complex frequency plane for several modes of an asymmetric ($L_1 \neq L_2$) microwave ring graph. This 3D plot shows another perspective of Fig. \ref{detS_complex}.}
\label{detS_complex_v2}
\end{figure*}

We show in Fig. \ref{detS_complex_v2} the reconstruction of complex $\det [S]$ over the complex frequency plane from a different perspective compared to Fig. \ref{detS_complex}, highlighting the phase variation in the region between the shape and Feshbach resonances.

\bibliography{TimeDelay&RingGraph.bib}










\end{document}